\documentclass[journal]{IEEEtran}

\usepackage{pdfpages} 

\usepackage{pgfplots}
\usepackage{pgfplotstable}

\usepackage{upgreek}
\usepackage{subfig}

\usepackage{color}
\usepackage{soul}
\usepackage{siunitx}

\usepackage{graphicx}
\usepackage{multirow}
\usepackage{siunitx}
\usepackage{multirow}

\newcolumntype{P}[1]{>{\centering\arraybackslash}p{#1}}
\newcolumntype{M}[1]{>{\centering\arraybackslash}m{#1}}

\usepackage{cite}

\makeatletter
\def\hlinewd#1{%
\noalign{\ifnum0=`}\fi\hrule \@height #1 %
\futurelet\reserved@a\@xhline}
\makeatother
\usepackage[tworuled,noline,linesnumbered]{algorithm2e}
\SetNlSkip{0.9em}
\SetNlSty{textbf}{L}{}

\usepackage{amssymb,amsmath}

\usepackage{amsmath}

\SetAlFnt{\footnotesize}
\SetAlCapFnt{\footnotesize}
\SetAlCapNameFnt{\footnotesize}

\usepackage{amssymb,amsmath}

\newenvironment{nospaceflalign}
 {\setlength{\abovedisplayskip}{3pt}\setlength{\belowdisplayskip}{3pt}%
  \csname flalign\endcsname}
 {\csname endflalign\endcsname\ignorespacesafterend}
\newcommand{\subparagraph}{}
\usepackage[compact]{titlesec}
\titlespacing{\section}{4pt}{4pt}{4pt}
\titlespacing{\subsection}{3pt}{3pt}{3pt}
\titlespacing{\subsubsection}{2pt}{2pt}{-8pt}

\usepackage{amsthm}

\newcommand{\setmuskip}[2]{#1=#2\relax}
\setmuskip{\medmuskip}{1mu plus 0mu minus 1mu}
\setmuskip{\thinmuskip}{1mu}
\setmuskip{\thickmuskip}{1mu plus 1mu}

\newcommand{\figurefontsize}{\footnotesize}

\newcommand{\equationfontsize}{\fontsize{9.5pt}{1}\selectfont}

\graphicspath{{./Fig/}}
\newcommand{\tum}{Technical University of Munich (TUM)}

\def\figname{Fig.}

\newcommand{\reviewhl}[1]{\textcolor{black}{#1}}

\newcommand{\titlenym}{VirtualSync+}
\newcommand{\papertitle}{VirtualSync+:
Timing Optimization with Virtual Synchronization
}

\begin{document}

\title{\papertitle}

\author{
Grace Li Zhang, Bing Li, Xing Huang, Xunzhao Yin, Cheng Zhuo,~\IEEEmembership{Senior Member, IEEE}, Masanori Hashimoto,~\IEEEmembership{Senior Member, IEEE}, Ulf Schlichtmann,~\IEEEmembership{Senior Member, IEEE}
\thanks{
This work was %
funded by the Deutsche Forschungsgemeinschaft (DFG, German Research Foundation)
-- Project Number 146371743 -- TRR 89 Invasive Computing. 

A preliminary version of this paper was published in
the Proceedings of Design Automation Conference (DAC), 2018\cite{Grace2018_DAC}. 
\textit{(Corresponding author: Xing Huang)}
}

        \thanks{Grace Li Zhang, Bing Li, Xing Huang and Ulf Schlichtmann are at the Chair of 
Electronic Design Automation,
        \tum, Munich 80333, Germany (e-mail: grace-li.zhang@tum.de; b.li@tum.de; xing.huang@tum.de;
ulf.schlichtmann@tum.de).}
        \thanks{Xunzhao Yin and Cheng Zhuo are with the 
College of Information Science and Electronic Engineering, Zhejiang University 
 (e-mail: xzyin1@zju.edu.cn; czhuo@zju.edu.cn).}
         \thanks{Masanori Hashimoto is with the
Department of Information Systems Engineering, Osaka University,
 (e-mail: hasimoto@ist.osaka-u.ac.jp).}

}

\maketitle
\markboth{IEEE TRANSACTIONS ON COMPUTER-AIDED DESIGN OF INTEGRATED CIRCUITS AND SYSTEMS}
{Zhang \MakeLowercase{\textit{et al.}}: \papertitle}

\IEEEpeerreviewmaketitle

\begin{abstract}

In digital circuit designs, sequential components such as flip-flops are used
to synchronize signal propagations. Logic
computations are aligned at and thus isolated by flip-flop stages.  
Although this fully synchronous style can reduce
design efforts significantly, it may affect circuit performance negatively,
because sequential components can only introduce delays into signal
propagations but never accelerate them.  In this paper, we propose a new
timing model, VirtualSync+, in which signals, specially those along critical
paths, are allowed to propagate through several sequential stages without
flip-flops.  Timing constraints are still satisfied at the boundary of the
optimized circuit to maintain a consistent interface with existing designs.
By removing clock-to-q delays and setup time requirements of flip-flops on critical
paths, the performance of a circuit can be pushed even beyond the limit of
traditional sequential designs. 
In addition, we further enhance the optimization with VirtualSync+ by fine-tuning with
commercial design tools, 
e.g., Design Compiler from
Synopsys, to 
achieve more accurate result.
Experimental results demonstrate that circuit
performance can be improved by up to 4\% (average 1.5\%) compared with that
after extreme retiming and sizing, while the increase of area is
still negligible. This timing performance is enhanced beyond 
the limit of traditional sequential designs. 
It also demonstrates that 
compared with those
after retiming and sizing, 
 the circuits with VirtualSync+ can achieve 
better timing performance under the same area cost 
or smaller area cost 
under the same clock period, respectively. 
 
\end{abstract}

\section{Introduction}

\IEEEPARstart{I}{n digital} circuit designs, clock frequency determines the timing performance
of %
circuits. In the traditional timing paradigm, sequential components,
e.g., edge-triggered flip-flops, synchronize signal propagations between pairs
of flip-flops.  Consequently, these propagations are blocked at flip-flops
until a clock edge arrives.  At an active clock edge, the data at the inputs 
of flip-flops are transferred to their outputs to drive the logic of the next stage.  
Therefore, combinational logic blocks are isolated by flip-flop stages. This fully synchronous
style can reduce design efforts significantly, since only 
timing constraints local to pairs of flip-flops need to be met.

Within the traditional timing paradigm, 
timing analysis and timing optimization 
have been explored extensively. 
\reviewhl{
In timing analysis, researchers focus on 
improving the execution efficiency and accuracy of analysis \cite{huiru2015,Cao2019,tsungwei2017,wire2020,ZhangLS16_iccad,ChenLS12}. 
In timing optimization, several methods have been proposed to improve 
timing performance. 
A widely adopted method is sizing, in which
logic gates are sized to improve objectives such
as clock frequency and area efficiency, while timing constraints between
flip-flops are satisfied.
Typical methods for gate sizing are based on Lagrangian Relaxation and sensitivity \cite{Wong1999,Hu2011,Jin2012,Huiru1999}.
The second method to improve circuit performance in the traditional paradigm
is retiming, which moves sequential components, e.g., 
flip-flops, but still preserves the correct functional behavior of circuits.
The existing retiming methods usually focus on reducing execution time while improving 
the performance of digital circuits \cite{Zhou2006,Aaron2008,Beerel2017,Singh2005,Huimei2021}.
}
Useful clock skew is the third method to enhance timing performance of digital circuits. 
For example, \cite{Grace18_test,Zhang201611,LICS111} 
intentionally exploits clock skews to improve the yield of digital circuits. 
Beyond these techniques, a further method to improve clock frequency is to introduce approximation in computational result \cite{Chuangtao2021}. 

Wave-pipelining is the third method to improve circuit performance, where
several logic waves are allowed to propagate through combinational paths without
intermediate sequential components.  
Wave-pipelining paths are different from multi-cycle paths where there is only one logic wave propagating 
along them at a moment. \reviewhl{
In timing constraints,  
multi-cycle paths only restrict their delays to be smaller than a specified upper bound \cite{Gupta_1994,Higuchi_2002} while 
wave-pipelining  paths restrict not only the upper bound but also the lower bound of their delays.}
Wave-pipelining provides a mechanism to make
the clock frequency of a circuit independent of the largest path delay, which
limits circuit performance in traditional circuit designs \cite{Burleson1998}.
As early as in \cite{wave93}, an algorithm to automatically equalize delays 
in combinational logic circuits is proposed to realize wave-pipelining.  
In \cite{Joy1993}, a linear method to minimize the clock period
using wave-pipelining is proposed. 
This method %
is also explored for majority-based
beyond-CMOS technologies to improve the throughput of \reviewhl{majority inverter graph
designs} in \cite{Micheli17}. 
Testing methods of wave-pipelined circuits are proposed in \cite{test94}.  
Recently, wave-pipelining 
is applied to accelerate the dot-product operations of neural network accelerators in \cite{Yehuda2021}. 
In addition, wave-pipelining has been applied to enhance netlist security\cite{ZLYPS18,Timingcamouflge+20}.

The first two methods above can be used separately or jointly to improve
circuit performance.  
However, sequential components are assumed to synchronize signal propagations
in these methods, where no signal propagation through sequential components is allowed 
except at the clock edges. 
This synchronization with sequential components achieves many benefits such
as reducing design efforts.  However, it limits circuit performance in two
regards. Firstly, sequential components have inherent clock-to-q delays and
impose setup time.  The former becomes a part of combinational
paths driven by the corresponding flip-flops and the latter requires a further 
part of the timing budget for the critical paths.  Secondly, delay imbalances
between flip-flop stages cannot be exploited since signal propagations are
blocked at flip-flops instead of being allowed to propagate through
flip-flops. Although clock skew scheduling can relieve this problem to some degree, 
it still suffers the inherent clock-to-q delays and setup time constraints of flip-flops. 
The third method above, wave-pipelining, allows signals to pass through
sequential stages without flip-flops. However, this technique %
is not compatible with
the traditional timing paradigm. 
Although a wave-pipelining utility that \reviewhl{interacts} with commercial tools is proposed in \cite{Yehuda2021} to achieve
delay balance, the proposed method in this work can only deal with generic combinational circuits without feedback loops, which 
restricts its application in sequential digital circuits.

In this paper, we propose a new timing model, \titlenym, which removes 
the confines of the traditional timing paradigm. Our contributions are as follows:
\begin{itemize} 
\newlength{\mylength} 
\setlength\mylength{\parskip}
\setlength\parskip{1.0mm}
\item 
In the proposed new timing model, sequential components and combinational 
logic gates are both considered as delay units. 
Combinational logic gates add linear delays of the same amount 
to short and long paths, where sequential components provide non-linear
delay effects, which provide different delay effects to fast and slow signal
propagations. %

\setlength\parskip{\mylength}

\item 
With the new timing model, a timing optimization framework is proposed to
allocate sequential components only at necessary locations in the circuit to
synchronize signal propagations, while the functionality of circuits is
maintained. The absence of flip-flops at some sequential stages allows 
a virtual synchronization to provide identical functionality as in the original circuit.
Consequently, the
original clock-to-q delays and setup requirements along the critical paths can
be removed to achieve a better circuit performance even beyond the limit of traditional
sequential designs.

\item
The optimization with \titlenym\ is further enhanced by fine-tuning with
commercial design tools,
e.g., Design Compiler from
Synopsys, to
achieve more accurate result. 
To achieve this fine-tuning, we first
optimize the circuits %
by reallocating sequential components.  
Afterwards,
the removal locations of flip-flops with respect to the circuits under optimization are extracted and
the corresponding wave-pipelining timing constraints compatible with commercial design tools are established.
These timing constraints are then incorporated into the optimization flow of commercial tools
to generate the optimized circuits. %

\end{itemize}
The rest of this paper is organized as follows. In
Section~\ref{sec:motivation}, we explain the motivation and the basic idea of
the proposed method. The timing optimization problem is formulated in Section~\ref{sec:problem_formulation}. 
In Section~\ref{sec:model}, we provide a detailed
description of the proposed timing model \titlenym. 
The proposed timing optimization %
with relaxed \titlenym\ timing model 
is described in Section~\ref{sec:impl}.  
Circuit fine-tuning with \titlenym\ in commercial tools is 
explained in Section~\ref{sec:integration}. 
Experimental results are reported in Section~\ref{sec:results}.  Conclusions
are drawn in Section~\ref{sec:conclusion}.

\section{Background and Motivation}\label{sec:motivation}

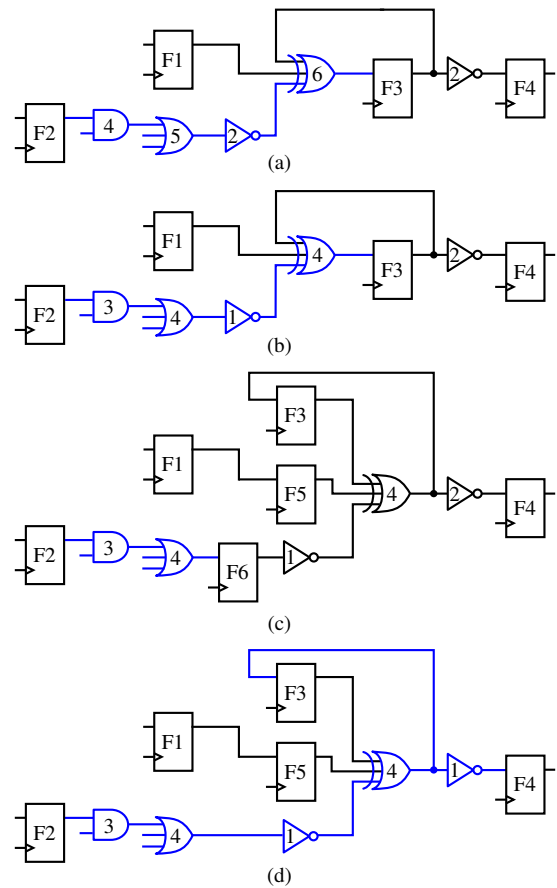
\begin{figure}[t]
{
\figurefontsize
\centering
\begingroup%
  \makeatletter%
  \providecommand\color[2][]{%
    \errmessage{(Inkscape) Color is used for the text in Inkscape, but the package 'color.sty' is not loaded}%
    \renewcommand\color[2][]{}%
  }%
  \providecommand\transparent[1]{%
    \errmessage{(Inkscape) Transparency is used (non-zero) for the text in Inkscape, but the package 'transparent.sty' is not loaded}%
    \renewcommand\transparent[1]{}%
  }%
  \providecommand\rotatebox[2]{#2}%
  \ifx\svgwidth\undefined%
    \setlength{\unitlength}{203.625bp}%
    \ifx\svgscale\undefined%
      \relax%
    \else%
      \setlength{\unitlength}{\unitlength * \real{\svgscale}}%
    \fi%
  \else%
    \setlength{\unitlength}{\svgwidth}%
  \fi%
  \global\let\svgwidth\undefined%
  \global\let\svgscale\undefined%
  \makeatother%
  \begin{picture}(1,1.62361754)%
    \put(0,0){\includegraphics[width=\unitlength]{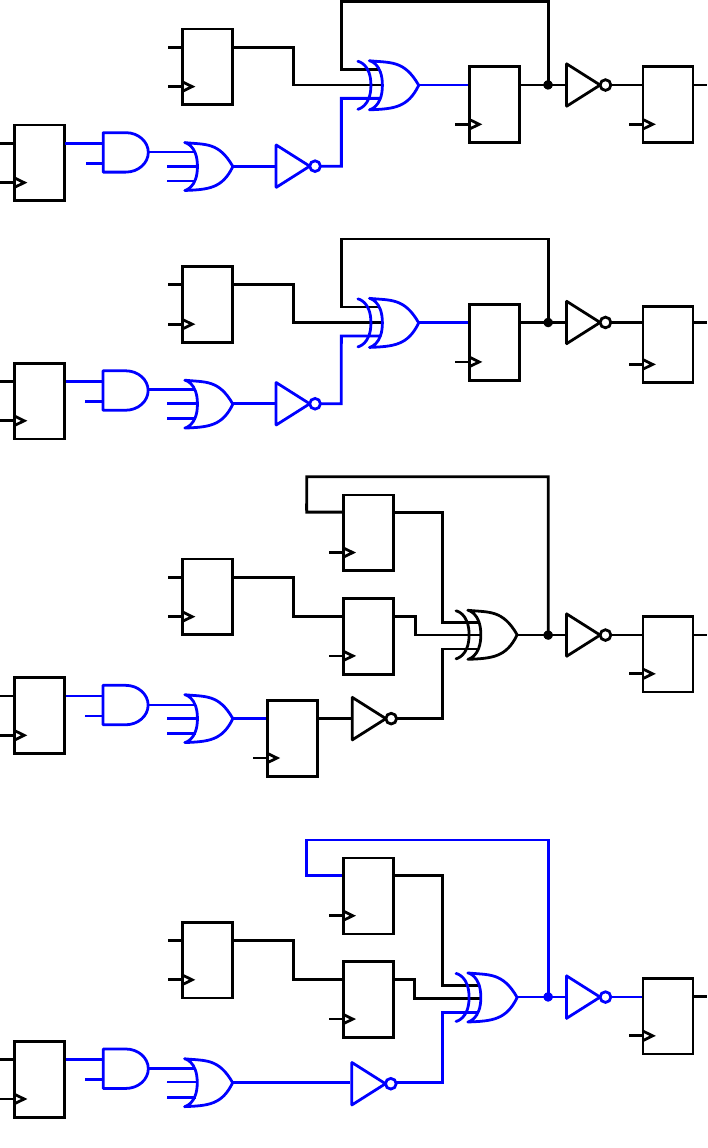}}%
    \put(0.4899986,0.47237928){\color[rgb]{0,0,0}\makebox(0,0)[b]{\smash{(c)}}}%
    \put(0.16442416,1.08448412){\color[rgb]{0,0,0}\makebox(0,0)[lt]{\begin{minipage}{0.09875972\unitlength}\raggedright 3\end{minipage}}}%
    \put(0.29265414,1.51646413){\color[rgb]{0,0,0}\makebox(0,0)[b]{\smash{F1}}}%
    \put(0.05508076,1.38064697){\color[rgb]{0,0,0}\makebox(0,0)[b]{\smash{F2}}}%
    \put(0.69857889,1.46317233){\color[rgb]{0,0,0}\makebox(0,0)[b]{\smash{F3}}}%
    \put(0.94404121,1.46317255){\color[rgb]{0,0,0}\makebox(0,0)[b]{\smash{F4}}}%
    \put(0.4899986,1.32668075){\color[rgb]{0,0,0}\makebox(0,0)[b]{\smash{(a)}}}%
    \put(0.4899986,0.98706862){\color[rgb]{0,0,0}\makebox(0,0)[b]{\smash{(b)}}}%
    \put(0.29265414,1.18041642){\color[rgb]{0,0,0}\makebox(0,0)[b]{\smash{F1}}}%
    \put(0.69857889,1.12712474){\color[rgb]{0,0,0}\makebox(0,0)[b]{\smash{F3}}}%
    \put(0.94404121,1.12356944){\color[rgb]{0,0,0}\makebox(0,0)[b]{\smash{F4}}}%
    \put(0.05508076,1.04384449){\color[rgb]{0,0,0}\makebox(0,0)[b]{\smash{F2}}}%
    \put(0.29265402,0.76680257){\color[rgb]{0,0,0}\makebox(0,0)[b]{\smash{F1}}}%
    \put(0.41258805,0.56684685){\color[rgb]{0,0,0}\makebox(0,0)[b]{\smash{F6}}}%
    \put(0.94404146,0.68594105){\color[rgb]{0,0,0}\makebox(0,0)[b]{\smash{F4}}}%
    \put(0.05508076,0.59894567){\color[rgb]{0,0,0}\makebox(0,0)[b]{\smash{F2}}}%
    \put(0.5200411,0.71128244){\color[rgb]{0,0,0}\makebox(0,0)[b]{\smash{F5}}}%
    \put(0.51988804,0.85745174){\color[rgb]{0,0,0}\makebox(0,0)[b]{\smash{F3}}}%
    \put(0.28796556,1.40120696){\color[rgb]{0,0,0}\makebox(0,0)[lt]{\begin{minipage}{0.09356184\unitlength}\raggedright 5\end{minipage}}}%
    \put(0.55053543,1.51692289){\color[rgb]{0,0,0}\makebox(0,0)[lt]{\begin{minipage}{0.09356184\unitlength}\raggedright 6\end{minipage}}}%
    \put(0.16452471,1.4204587){\color[rgb]{0,0,0}\makebox(0,0)[lt]{\begin{minipage}{0.09875972\unitlength}\raggedright 4\end{minipage}}}%
    \put(0.28833104,0.62062956){\color[rgb]{0,0,0}\makebox(0,0)[lt]{\begin{minipage}{0.09356184\unitlength}\raggedright 4\end{minipage}}}%
    \put(0.55072972,1.18071737){\color[rgb]{0,0,0}\makebox(0,0)[lt]{\begin{minipage}{0.09356184\unitlength}\raggedright 4\end{minipage}}}%
    \put(0.16444729,0.63856925){\color[rgb]{0,0,0}\makebox(0,0)[lt]{\begin{minipage}{0.09875972\unitlength}\raggedright 3\end{minipage}}}%
    \put(0.68835362,0.73910995){\color[rgb]{0,0,0}\makebox(0,0)[lt]{\begin{minipage}{0.09356184\unitlength}\raggedright 4\end{minipage}}}%
    \put(0.28815986,1.06466675){\color[rgb]{0,0,0}\makebox(0,0)[lt]{\begin{minipage}{0.09356184\unitlength}\raggedright 4\end{minipage}}}%
    \put(0.4899986,0.00657466){\color[rgb]{0,0,0}\makebox(0,0)[b]{\smash{(d)}}}%
    \put(0.29265402,0.25314178){\color[rgb]{0,0,0}\makebox(0,0)[b]{\smash{F1}}}%
    \put(0.94404121,0.17391536){\color[rgb]{0,0,0}\makebox(0,0)[b]{\smash{F4}}}%
    \put(0.05508089,0.08459245){\color[rgb]{0,0,0}\makebox(0,0)[b]{\smash{F2}}}%
    \put(0.51988732,0.34379167){\color[rgb]{0,0,0}\makebox(0,0)[b]{\smash{F3}}}%
    \put(0.28803945,0.10570886){\color[rgb]{0,0,0}\makebox(0,0)[lt]{\begin{minipage}{0.09356184\unitlength}\raggedright 4\end{minipage}}}%
    \put(0.16467606,0.12411261){\color[rgb]{0,0,0}\makebox(0,0)[lt]{\begin{minipage}{0.09875972\unitlength}\raggedright 3\end{minipage}}}%
    \put(0.68835359,0.22618361){\color[rgb]{0,0,0}\makebox(0,0)[lt]{\begin{minipage}{0.09356184\unitlength}\raggedright 4\end{minipage}}}%
    \put(0.80421214,1.51703416){\color[rgb]{0,0,0}\makebox(0,0)[lt]{\begin{minipage}{0.08103157\unitlength}\raggedright 2\end{minipage}}}%
    \put(0.5198833,0.19762261){\color[rgb]{0,0,0}\makebox(0,0)[b]{\smash{F5}}}%
    \put(0.3934767,1.40228971){\color[rgb]{0,0,0}\makebox(0,0)[lt]{\begin{minipage}{0.08103157\unitlength}\raggedright 2\end{minipage}}}%
    \put(0.80421214,0.73908933){\color[rgb]{0,0,0}\makebox(0,0)[lt]{\begin{minipage}{0.08103157\unitlength}\raggedright 2\end{minipage}}}%
    \put(0.80421214,1.18122679){\color[rgb]{0,0,0}\makebox(0,0)[lt]{\begin{minipage}{0.08103157\unitlength}\raggedright 2\end{minipage}}}%
    \put(0.39347672,1.06624197){\color[rgb]{0,0,0}\makebox(0,0)[lt]{\begin{minipage}{0.08103157\unitlength}\raggedright 1\end{minipage}}}%
    \put(0.50117049,0.62091221){\color[rgb]{0,0,0}\makebox(0,0)[lt]{\begin{minipage}{0.08103157\unitlength}\raggedright 1\end{minipage}}}%
    \put(0.50061971,0.10453625){\color[rgb]{0,0,0}\makebox(0,0)[lt]{\begin{minipage}{0.08103157\unitlength}\raggedright 1\end{minipage}}}%
    \put(0.80421214,0.22706395){\color[rgb]{0,0,0}\makebox(0,0)[lt]{\begin{minipage}{0.08103157\unitlength}\raggedright 1\end{minipage}}}%
  \end{picture}%
\endgroup%

\caption{Timing optimization methods. Delays of logic gates are shown on the gates. The clock-to-q delay ($t_{cq}$), setup time ($t_{su}$) and hold time ($t_{h}$) of a flip-flop are 3, 1 and 1, respectively. (a) Original circuit. (b) Sized circuit. (c) Circuit after retiming.  (d) Circuit after optimization using \titlenym.}
\label{fig:concept}
}
\end{figure}
In traditional digital circuits, sequential components such as flip-flops
synchronize signal propagations between pairs of flip-flops using a global
clock signal, as shown in \figname~\ref{fig:concept}(a). The combinational
path between F2 and F3 is critical with a path delay equal to 17. 
Assume that the clock-to-q delay, the
setup time and the hold time of a flip-flop are 3, 1, and 1, respectively. The
minimum clock period of this circuit is thus equal to 21. 

To reduce the clock period, logic gates with smaller delays can be selected
from the library to accelerate signal propagations on the critical paths of the circuit, at the cost
of additional area overhead, leading to the circuit shown in
\figname~\ref{fig:concept}(b), where the logic gates that are not on the
critical path still have their original delays for the sake of saving area.
After sizing, the minimum clock period of this circuit is reduced to 16 units.
To reduce the clock period further, retiming can be deployed to move F3 to the
left of the XOR gate as shown in \figname~\ref{fig:concept}(c), leading to a
minimum clock period equal to 11.

The circuit in \figname~\ref{fig:concept}(c) has reached the limit of timing 
performance in the traditional timing model, and no other method except a logic
redesign can reduce the clock period further. 
However, this strict timing constraint can still be relaxed by removing F6
from the circuit, leading to the circuit in \figname~\ref{fig:concept}(d). 
If the signal from from F2 can reach the sink flip-flops F3 and F4 after the
next rising clock edge and before the rising edge two periods later, data can
still be latched by F3 and F4 correctly. Since the inverter before F4 can also
be sized further, \reviewhl{the largest path delay including clock-to-q delay is 16. 
The minimum clock period constrained by this path is thus 
(16+1)/2=8.5. In this scenario, 
the minimum clock period of the circuit is limited by the delay of the path from F5 to F4, which is 9,  18.2\% lower than retiming.}
\reviewhl{The circuit in \figname~\ref{fig:concept}(d) is one of the solutions with \titlenym, where F6 is removed and F5 as well as F3 are inserted back to block fast signals.}

Since F6 can be removed from the circuit without affecting its function in fact, it
makes no contribution to the logic function or timing performance in
\figname~\ref{fig:concept}(c).  However, the flip-flop F5 in
\figname~\ref{fig:concept}(c) cannot be removed, because the signal from F1
should also arrive at F4 later than one clock period.
Without F5, the signal from F1 arrives at F4 even before the next rising clock edge, 
 and thus a loss of logic synchronization arises compared with the circuit in \figname~\ref{fig:concept}(a).
Comparing \figname~\ref{fig:concept}(b) and
\figname~\ref{fig:concept}(d), \reviewhl{we can see that F3 in
\figname~\ref{fig:concept}(b) blocks the fast path from F1 to F4 and breaks the feedback loop to avoid loss of 
logic synchronization}, but it degrades the circuit performance by delaying
the signal from F2 to F4 too.   

The concept to allow logic signals to span several sequential stages without a
flip-flop separating them is called \textit{wave-pipelining}
\cite{Burleson1998}. Previously, this technique has only been explored in the
context of circuit design, where the numbers of waves on logic paths should be
defined and their synchronization should be maintained by designers during the
design phase. Since logic design and timing cannot be handled separately as in traditional synchronous designs, 
wave-pipelining becomes incompatible with the traditional fully
synchronous design paradigm, which prevents its adoption in practical designs.
In \titlenym, we introduce a new timing model that allows multiple waves on
logic paths as a technique of timing optimization for sequential digital circuits in the
traditional design style. The resulting circuits still provide correct timing
interfaces to sequential components, e.g., flip-flops, at the boundary of the
optimized circuits to maintain timing compatibility.

\section{Problem Formulation}\label{sec:problem_formulation}

In digital circuits, \reviewhl{we propose that the essential function of sequential components is to
delay signals along fast paths in a circuit.} For example,  in
\figname~\ref{fig:concept}(d), F5 must be kept in the circuit to delay the
signal propagation from F1 to F4. The sequential components that only sit on
the critical path can thus be removed to improve circuit performance, such as F6 in
\figname~\ref{fig:concept}(d).

In the \titlenym\ framework, we 
remove all flip-flops and then identify the necessary locations
to block fast signals using combinational gates and sequential components, e.g., buffers, flip-flops,
and latches. The advantage of this formulation is that it is possible to
insert the minimum number of delay units into the circuit to achieve the theoretical
minimum clock period.

The problem formulation of \titlenym\ is described as follows:

\noindent \textit{Given}: the netlist of a digital circuit; the delay
information of the circuit; the target clock period T.

\noindent \textit{Output}: a circuit with adjusted number and locations of sequential
components; logic gates with new sizes; inserted delay units, e.g., buffers.

\noindent \textit{Objectives}: the circuit should maintain the same function viewed
from the sequential components at the boundary of the optimized circuit; the
target timing specification should be met; the area of the optimized circuit 
should be reduced. 

\reviewhl{In the following sections, we will introduce the proposed \titlenym\ timing optimization framework. 
Delay units and their insertion with a complete timing model %
are first explained in Section~\ref{sec:model}. 
Since it is time-consuming to optimize circuits with the complete model,
we introduce a heuristic optimization framework with relaxed \titlenym\ timing model %
in Section~\ref{sec:impl}. 
This optimization is further applied together with %
Design Compiler from Synopsys to achieve more accurate results
in Section~\ref{sec:integration}.}

\section{\titlenym\  Timing Model}\label{sec:model}
  
\subsection{Delay Units} \label{sec:delay_units}

In the \titlenym\ framework, we first remove all sequential components, 
flip-flops, from the circuit under optimization.  
Consequently, logic synchronization may be lost because  
signals across fast paths may arrive at flip-flops in incorrect clock cycles, e.g., 
earlier than specified, or timing violations may be incurred.
In addition, signals along combinational loops should also be blocked to avoid
the loss of logic synchronization. For example, in  
\figname~\ref{fig:concept}(d), the combinational loop across the
XOR gate must have
a sequential component; otherwise a signal loses synchronization after traveling
across it many times. 

\begin{figure}[t]
{
\figurefontsize
\centering
\begingroup%
  \makeatletter%
  \providecommand\color[2][]{%
    \errmessage{(Inkscape) Color is used for the text in Inkscape, but the package 'color.sty' is not loaded}%
    \renewcommand\color[2][]{}%
  }%
  \providecommand\transparent[1]{%
    \errmessage{(Inkscape) Transparency is used (non-zero) for the text in Inkscape, but the package 'transparent.sty' is not loaded}%
    \renewcommand\transparent[1]{}%
  }%
  \providecommand\rotatebox[2]{#2}%
  \ifx\svgwidth\undefined%
    \setlength{\unitlength}{266.83523696bp}%
    \ifx\svgscale\undefined%
      \relax%
    \else%
      \setlength{\unitlength}{\unitlength * \real{\svgscale}}%
    \fi%
  \else%
    \setlength{\unitlength}{\svgwidth}%
  \fi%
  \global\let\svgwidth\undefined%
  \global\let\svgscale\undefined%
  \makeatother%
  \begin{picture}(1,0.41493637)%
    \put(0.57861055,0.13877297){\color[rgb]{0,0,0}\makebox(0,0)[lb]{\smash{$D*T$+$t_{cq}$}}}%
    \put(0.26817143,0.21667492){\color[rgb]{0,0,0}\makebox(0,0)[lb]{\smash{$T$+$t_{cq}$}}}%
    \put(0.64751938,0.3828556){\color[rgb]{0,0,0}\makebox(0,0)[lb]{\smash{$s_u$}}}%
    \put(0.30154342,0.3828556){\color[rgb]{0,0,0}\makebox(0,0)[lb]{\smash{$s_u$}}}%
    \put(0.02570067,0.3828556){\color[rgb]{0,0,0}\makebox(0,0)[rb]{\smash{$s_u$}}}%
    \put(0,0){\includegraphics[width=\unitlength,page=1]{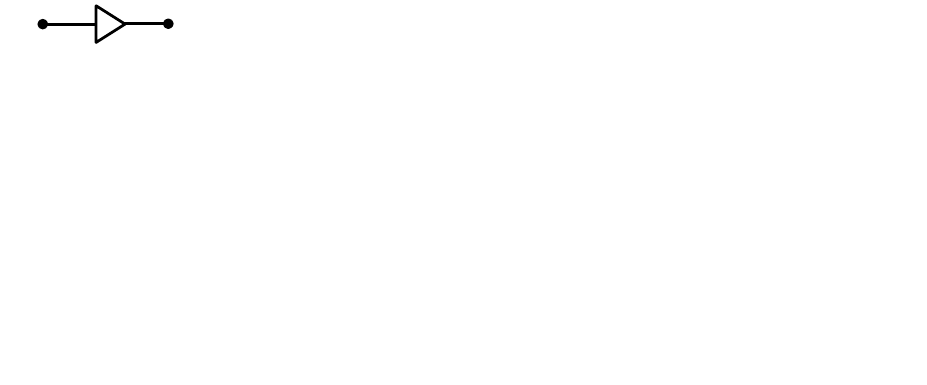}}%
    \put(0.20104917,0.3828556){\color[rgb]{0,0,0}\makebox(0,0)[lb]{\smash{$s_v$}}}%
    \put(-0.02870839,0.36110503){\color[rgb]{0,0,0}\makebox(0,0)[lt]{\begin{minipage}{0.2960818\unitlength}\centering $t_{d}$\end{minipage}}}%
    \put(0,0){\includegraphics[width=\unitlength,page=2]{difference.pdf}}%
    \put(0.44730036,0.35560225){\color[rgb]{0,0,0}\makebox(0,0)[b]{\smash{F}}}%
    \put(0,0){\includegraphics[width=\unitlength,page=3]{difference.pdf}}%
    \put(0.5388584,0.3828556){\color[rgb]{0,0,0}\makebox(0,0)[lb]{\smash{$s_v$}}}%
    \put(0,0){\includegraphics[width=\unitlength,page=4]{difference.pdf}}%
    \put(0.53160587,0.05294032){\color[rgb]{0,0,0}\makebox(0,0)[lb]{\smash{$s_u$}}}%
    \put(0.31769981,0.25995475){\color[rgb]{0,0,0}\makebox(0,0)[lb]{\smash{$s_v$}}}%
    \put(0,0){\includegraphics[width=\unitlength,page=5]{difference.pdf}}%
    \put(0.50704041,0.04331584){\color[rgb]{0,0,0}\makebox(0,0)[b]{\smash{$T$}}}%
    \put(0,0){\includegraphics[width=\unitlength,page=6]{difference.pdf}}%
    \put(0.78938215,0.3556022){\color[rgb]{0,0,0}\makebox(0,0)[b]{\smash{L}}}%
    \put(0,0){\includegraphics[width=\unitlength,page=7]{difference.pdf}}%
    \put(0.88233903,0.3828556){\color[rgb]{0,0,0}\makebox(0,0)[lb]{\smash{$s_v$}}}%
    \put(0,0){\includegraphics[width=\unitlength,page=8]{difference.pdf}}%
    \put(0.85746879,0.0441166){\color[rgb]{0,0,0}\makebox(0,0)[b]{\smash{$T$}}}%
    \put(0,0){\includegraphics[width=\unitlength,page=9]{difference.pdf}}%
    \put(0.79076151,0.0441166){\color[rgb]{0,0,0}\makebox(0,0)[b]{\smash{$D*T$}}}%
    \put(0,0){\includegraphics[width=\unitlength,page=10]{difference.pdf}}%
    \put(0.13470042,0.02439469){\color[rgb]{0,0,0}\makebox(0,0)[lt]{\begin{minipage}{0.10493367\unitlength}\raggedright (a)\end{minipage}}}%
    \put(0.43357978,0.02438463){\color[rgb]{0,0,0}\makebox(0,0)[lt]{\begin{minipage}{0.10493367\unitlength}\raggedright (b)\end{minipage}}}%
    \put(0.78477075,0.02439721){\color[rgb]{0,0,0}\makebox(0,0)[lt]{\begin{minipage}{0.16489576\unitlength}\raggedright (c)\end{minipage}}}%
    \put(0,0){\includegraphics[width=\unitlength,page=11]{difference.pdf}}%
    \put(0.49764452,0.26995343){\color[rgb]{0,0,0}\makebox(0,0)[b]{\smash{$t_{su}$}}}%
    \put(0,0){\includegraphics[width=\unitlength,page=12]{difference.pdf}}%
    \put(0.23246719,0.05294023){\color[rgb]{0,0,0}\makebox(0,0)[lb]{\smash{$s_u$}}}%
    \put(0.05165726,0.25995466){\color[rgb]{0,0,0}\makebox(0,0)[rb]{\smash{$s_v$}}}%
    \put(0,0){\includegraphics[width=\unitlength,page=13]{difference.pdf}}%
    \put(0.90211327,0.05294032){\color[rgb]{0,0,0}\makebox(0,0)[b]{\smash{$s_u$}}}%
    \put(0.66853357,0.25995475){\color[rgb]{0,0,0}\makebox(0,0)[lb]{\smash{$s_v$}}}%
    \put(-0.24707104,0.1246798){\color[rgb]{0,0,0}\makebox(0,0)[lt]{\begin{minipage}{0.2960818\unitlength}\raggedleft $t_{d}$\end{minipage}}}%
    \put(0,0){\includegraphics[width=\unitlength,page=14]{difference.pdf}}%
    \put(0.84807291,0.26995343){\color[rgb]{0,0,0}\makebox(0,0)[b]{\smash{$t_{su}$}}}%
    \put(0,0){\includegraphics[width=\unitlength,page=15]{difference.pdf}}%
    \put(0.13852444,0.05003522){\color[rgb]{0,0,0}\makebox(0,0)[b]{\smash{input  time }}}%
    \put(0.02886076,0.18743926){\color[rgb]{0,0,0}\rotatebox{-90}{\makebox(0,0)[b]{\smash{output }}}}%
    \put(0.00005307,0.18713188){\color[rgb]{0,0,0}\rotatebox{-90}{\makebox(0,0)[b]{\smash{time}}}}%
    \put(0,0){\includegraphics[width=\unitlength,page=16]{difference.pdf}}%
    \put(0.4082514,0.10365585){\color[rgb]{0,0,0}\makebox(0,0)[b]{\smash{$t_{h}$}}}%
    \put(0,0){\includegraphics[width=\unitlength,page=17]{difference.pdf}}%
    \put(0.75394296,0.11419767){\color[rgb]{0,0,0}\makebox(0,0)[b]{\smash{$t_{h}$}}}%
    \put(0,0){\includegraphics[width=\unitlength,page=18]{difference.pdf}}%
    \put(0.62658279,0.21407642){\color[rgb]{0,0,0}\makebox(0,0)[b]{\smash{output }}}%
    \put(0.62668526,0.18793789){\color[rgb]{0,0,0}\makebox(0,0)[b]{\smash{time}}}%
    \put(0.77797574,0.26505482){\color[rgb]{0,0,0}\makebox(0,0)[b]{\smash{input }}}%
    \put(0.77512283,0.23891628){\color[rgb]{0,0,0}\makebox(0,0)[b]{\smash{time}}}%
    \put(0,0){\includegraphics[width=\unitlength,page=19]{difference.pdf}}%
    \put(0.35213095,0.06404219){\color[rgb]{0,0,0}\makebox(0,0)[lt]{\begin{minipage}{0.03759505\unitlength}\raggedright 0\end{minipage}}}%
    \put(0.70376127,0.06213979){\color[rgb]{0,0,0}\makebox(0,0)[lt]{\begin{minipage}{0.03759505\unitlength}\raggedright 0\end{minipage}}}%
    \put(0,0){\includegraphics[width=\unitlength,page=20]{difference.pdf}}%
  \end{picture}%
\endgroup%

\caption{Properties of delay units, assuming the launching and capturing times of an active clock edge are 0 and $T$, respectively. (a) Linear delaying effect of a combinational delay unit. (b) Constant delaying effect of a flip-flop. (c) Piecewise delaying effect of a latch.}
\label{fig:difference}
}
\end{figure}

To slow down a signal, three different components can be used as
delay units, namely, combinational gates such as buffers, 
flip-flops, and latches, which exhibit different delay
characteristics, as shown in \figname~\ref{fig:difference}, 
\reviewhl{where input/output time refers to the input/out arrival time of a signal}.

In \figname~\ref{fig:difference}(a), a combinational delay unit adds the same amount
of delay to any input signal.
Consequently, the arrival time $s_v$ at the output of the combinational delay 
unit is linear to the arrival time $s_u$ 
at the input of the
delay unit. Therefore, the absolute gap between \reviewhl{the early and late arrival times of signals} through
short and long paths does not change when a combinational delay unit is passed through.

In delaying input signals, a flip-flop, as a sequential delay unit, behaves completely differently from a
combinational delay unit, as shown in \figname~\ref{fig:difference}(b). If the
arrival time of a signal falls into the time window [$t_h$, $T-t_{su}$], where
$t_h$ is the hold time and $t_{su}$ is the setup time, the output signal
always leaves at the time $T+t_{cq}$, with $t_{cq}$ as the clock-to-q delay of
the flip-flop. Therefore, the gap between \reviewhl{the early and late arrival times of two signals} 
reaching the input of a flip-flop is always reduced to zero at the output of the flip-flop.
This is a very useful property because the delays of short paths and long
paths in a circuit may differ significantly after all sequential components
are removed from the circuit under optimization.
 For many short paths, it is not possible to increase their delays
by adding combinational delay units such as buffers to them, because the combinational delay units on
the short paths may also appear on other long paths.
The increased delays along long paths might affect circuit performance negatively.
 Flip-flops are thus of
great use in this scenario, because short paths receive more delay
padding than long paths to align logic waves in the circuit.  

As the second type of sequential delay units, level-sensitive latches have a delay property combining those of combinational
delay units and flip-flops, as shown in \figname~\ref{fig:difference}(c), where
$0<D<1$ is the duty cycle of the clock signal.  Assume that a latch is
non-transparent in the first part of the clock period and transparent in the
second part of the clock period. If two input signals arrive at a latch when
it is non-transparent, the output gap is reduced to zero. If both signals
arrive at a latch when it is transparent, the gap remains unchanged. However,
if the fast signal reaches the latch when it is non-transparent while the slow
signal reaches it when it is transparent, \reviewhl{the gap between the early and late arrival times 
of two signals is} %
neither zero nor unchanged.  Instead, it takes a value between the two extreme
cases as illustrated in \figname~\ref{fig:difference}(c).  This property gives
us more flexibility to modulate signals with different arrival times, specifically those along 
critical paths where fast signals require more delay padding and
slow signals should not be affected.

\subsection{Relative Timing References} \label{sec:relative_timing}
In \figname~\ref{fig:concept}(a), if all the logic gates
and flip-flop F3 are considered as the circuit under optimization, 
F1, F2 and F4 are thus the boundary flip-flops. 
No matter how signals inside the circuit propagate, 
the function of the whole circuit is still maintained
if we can guarantee that for any input pattern at flip-flops F1 and
F2 the circuit produces the same result at F4 at the same clock cycle as the original circuit. 

\begin{figure}[t]
{
\figurefontsize
\centering
\begingroup%
  \makeatletter%
  \providecommand\color[2][]{%
    \errmessage{(Inkscape) Color is used for the text in Inkscape, but the package 'color.sty' is not loaded}%
    \renewcommand\color[2][]{}%
  }%
  \providecommand\transparent[1]{%
    \errmessage{(Inkscape) Transparency is used (non-zero) for the text in Inkscape, but the package 'transparent.sty' is not loaded}%
    \renewcommand\transparent[1]{}%
  }%
  \providecommand\rotatebox[2]{#2}%
  \ifx\svgwidth\undefined%
    \setlength{\unitlength}{231.41862825bp}%
    \ifx\svgscale\undefined%
      \relax%
    \else%
      \setlength{\unitlength}{\unitlength * \real{\svgscale}}%
    \fi%
  \else%
    \setlength{\unitlength}{\svgwidth}%
  \fi%
  \global\let\svgwidth\undefined%
  \global\let\svgscale\undefined%
  \makeatother%
  \begin{picture}(1,0.20733316)%
    \put(0,0){\includegraphics[width=\unitlength,page=1]{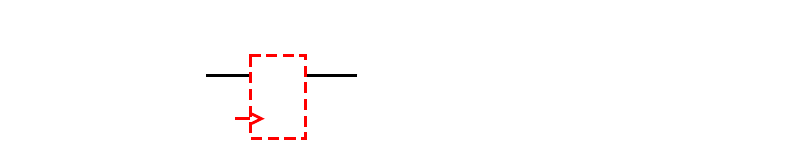}}%
    \put(0.34616614,0.07447539){\color[rgb]{1,0,0}\makebox(0,0)[b]{\smash{F2}}}%
    \put(0,0){\includegraphics[width=\unitlength,page=2]{minus_T.pdf}}%
    \put(0.0545465,0.07447539){\color[rgb]{0,0,0}\makebox(0,0)[b]{\smash{F1}}}%
    \put(0,0){\includegraphics[width=\unitlength,page=3]{minus_T.pdf}}%
    \put(0.64994743,0.07447549){\color[rgb]{0,0.50196078,0}\makebox(0,0)[b]{\smash{F3}}}%
    \put(0,0){\includegraphics[width=\unitlength,page=4]{minus_T.pdf}}%
    \put(0.28893854,0.13712383){\color[rgb]{0,0,0}\makebox(0,0)[b]{\smash{u}}}%
    \put(0.40641184,0.13590153){\color[rgb]{0,0,0}\makebox(0,0)[b]{\smash{v}}}%
    \put(0.58866169,0.13590153){\color[rgb]{0,0,0}\makebox(0,0)[b]{\smash{w}}}%
    \put(0.70872893,0.13610033){\color[rgb]{0,0,0}\makebox(0,0)[b]{\smash{t}}}%
    \put(0,0){\includegraphics[width=\unitlength,page=5]{minus_T.pdf}}%
    \put(0.94459541,0.07215795){\color[rgb]{0,0,0}\makebox(0,0)[b]{\smash{F4}}}%
    \put(0.93086125,-0.15308941){\color[rgb]{0,0,0}\makebox(0,0)[lt]{\begin{minipage}{0.08642343\unitlength}\raggedright \end{minipage}}}%
    \put(0.20087959,0.09943486){\color[rgb]{0,0,0}\makebox(0,0)[b]{\smash{11}}}%
    \put(0.50024485,0.09929243){\color[rgb]{0,0,0}\makebox(0,0)[b]{\smash{3}}}%
    \put(0.79988135,0.09913943){\color[rgb]{0,0,0}\makebox(0,0)[b]{\smash{2}}}%
    \put(0.3465557,0.00030594){\color[rgb]{0,0,0}\makebox(0,0)[b]{\smash{-10}}}%
    \put(0.65031506,0.00030594){\color[rgb]{0,0,0}\makebox(0,0)[b]{\smash{-10}}}%
    \put(0.28894967,0.19091776){\color[rgb]{0,0,0}\makebox(0,0)[b]{\smash{$s_u$=14}}}%
    \put(0.4064473,0.19091776){\color[rgb]{0,0,0}\makebox(0,0)[b]{\smash{$s_v$=4}}}%
    \put(0.58847135,0.19091776){\color[rgb]{0,0,0}\makebox(0,0)[b]{\smash{$s_w$=7}}}%
    \put(0.70869976,0.19091776){\color[rgb]{0,0,0}\makebox(0,0)[b]{\smash{$s_t$=3}}}%
    \put(0.888135,0.13610033){\color[rgb]{0,0,0}\makebox(0,0)[b]{\smash{z}}}%
    \put(0,0){\includegraphics[width=\unitlength,page=6]{minus_T.pdf}}%
    \put(0.11239238,0.13712383){\color[rgb]{0,0,0}\makebox(0,0)[b]{\smash{o}}}%
    \put(0.88814736,0.19091776){\color[rgb]{0,0,0}\makebox(0,0)[b]{\smash{$s_z$=5}}}%
    \put(0.11232436,0.19091776){\color[rgb]{0,0,0}\makebox(0,0)[b]{\smash{$s_o$=3}}}%
    \put(-0.12587701,0.01368065){\color[rgb]{0,0,0}\makebox(0,0)[lt]{\begin{minipage}{0.35933042\unitlength}\centering boundary  \end{minipage}}}%
    \put(0.765025,0.01368065){\color[rgb]{0,0,0}\makebox(0,0)[lt]{\begin{minipage}{0.35933042\unitlength}\centering boundary\end{minipage}}}%
  \end{picture}%
\endgroup%

\caption{Concept of relative timing references. Clock period T=10. Clock-to-q delay $t_{cq}$=3. Both setup time $t_{su}$ and hold time $t_h$ are equal to 1. F3 is kept in the optimized circuit and F2 is not included.}
\label{fig:minus_T}
}
\end{figure}

Consider a general case in \figname~\ref{fig:minus_T}, where F1 and F4 are the
boundary flip-flops and F2 and F3 are removed in the initial circuit for
optimization. At F4, the arrival times are required to meet the setup and hold
time constraints, written as
\begin{align}
s_z+t_{su}\le T \label{eq:simple_setup}\\
s^\prime_z\ge t_{h}\label {eq:simple_hold}\
\end{align}
where $s_z$ and $s^\prime_z$ are the latest and earliest arrival times at $z$.
These two constraints in fact are defined with respect to the
rising clock edge at F3, since the clock period T in (\ref{eq:simple_setup})
shows that the signal should arrive at F4 within one clock period.  Although
F2 and F3 are removed from the circuit, the constraints at F4 should still be the 
same as (\ref{eq:simple_setup})-(\ref{eq:simple_hold}) to maintain the
compatibility of the timing interface at the boundary flip-flops.

In the general case in \figname~\ref{fig:minus_T},
we can also observe that the timing constraint at F3 in
the original circuit is also defined with respect to the rising clock edge at
F2. This definition can be chained further back until the source flip-flop F1
at the boundary is be reached. We call the locations of these removed
flip-flops such as F2 and F3 
\textbf{\textit{anchor points}}.
After all sequential components are removed from
the circuit under optimization, these anchor points still allow to relate timing information to boundary 
flip-flops. Every time when a signal passes an anchor point, its arrival time
is converted by subtracting $T$ in \titlenym. When a signal finally arrives at a boundary
flip-flop along a combinational path, its arrival time must be converted 
so many times as the number of flip-flops on the path,
so that
(\ref{eq:simple_setup})-(\ref{eq:simple_hold}) is still valid.

In \figname~\ref{fig:minus_T}, assume that F2 is removed but F3 is
inserted back in the optimized circuit. The arrival time $s_u$ is subtracted
by the clock period T=10 to convert it with respect to the time at F1,
leading to $s_v$=4. 
The arrival time $s_w$ is defined with
respect to the previous flip-flop before F3, so that the timing constraints
can be checked using (\ref{eq:simple_setup})-(\ref{eq:simple_hold}). 
Since the arrival time before F4 should meet
its timing constraints, F3 thus cannot be removed. Otherwise, the arrival time
$s_t$ would be equal to 7-10=-3. Accordingly, the arrival time $s_z$
becomes -3+2=-1, definitely violating the hold time constraint in
(\ref{eq:simple_hold}).

Since F3 is kept in the optimized circuit, it introduces the delay with the
property shown in \figname~\ref{fig:difference}(b). 
The arrival time after this sequential delay unit thus becomes
$T+t_{cq}$=13. This signal at $t$ in \figname~\ref{fig:minus_T} also passes an anchor point. Therefore, the arrival
time $s_t$ is equal to 3, leading to no timing violation at F4. This example
demonstrates that the timing constraints at the boundary flip-flops force the
usage of the internal sequential delay units. \reviewhl{The model to automatically insert these
delay units} will be explained in the next section.

\subsection{Synchronizing Logic Waves by Delay Units} \label{sec:mode_delay_unit}

With 
all flip-flops removed
from the circuit under optimization, 
we only need to delay signals that are so fast that
they reach boundary flip-flops too early; signals that propagate 
slowly are already on the critical paths, thus requiring no additional
delay. Since it is not straightforward to determine the locations for inserting
additional delays, we formulate this task as an ILP problem and solve it
later with introduced heuristic steps. The values of variables in the following sections are 
determined by the solver, unless they are declared as constants explicitly.

The scenario of delay insertion at a circuit node, i.e., a logic gate, is illustrated in 
\figname~\ref{fig:circuit_graph}, where a combinational delay unit $\xi_{uv}$ may be
inserted, the original delay of the logic gate may be sized, and a sequential
delay unit may be inserted to block fast and slow signals with different
delays. Furthermore, the number of flip-flops \reviewhl{between $t$ and $z$ in the original circuit} is 
represented by an integer constant $\lambda_{tz}$. When $\lambda_{tz}$$\ge$1, an anchor point is found at the location between $t$ and $z$. 
$\lambda_{tz}$ is used to convert arrival times.

\subsubsection{Combinational delay unit and gate sizing}\label{sec:comb}
\begin{figure}[t]
{%
\centering
\begingroup%
  \makeatletter%
  \providecommand\color[2][]{%
    \errmessage{(Inkscape) Color is used for the text in Inkscape, but the package 'color.sty' is not loaded}%
    \renewcommand\color[2][]{}%
  }%
  \providecommand\transparent[1]{%
    \errmessage{(Inkscape) Transparency is used (non-zero) for the text in Inkscape, but the package 'transparent.sty' is not loaded}%
    \renewcommand\transparent[1]{}%
  }%
  \providecommand\rotatebox[2]{#2}%
  \ifx\svgwidth\undefined%
    \setlength{\unitlength}{161.1830755bp}%
    \ifx\svgscale\undefined%
      \relax%
    \else%
      \setlength{\unitlength}{\unitlength * \real{\svgscale}}%
    \fi%
  \else%
    \setlength{\unitlength}{\svgwidth}%
  \fi%
  \global\let\svgwidth\undefined%
  \global\let\svgscale\undefined%
  \makeatother%
  \begin{picture}(1,0.38419077)%
    \put(0,0){\includegraphics[width=\unitlength,page=1]{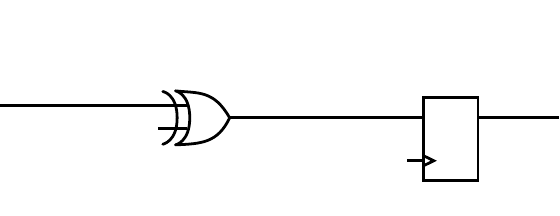}}%
    \put(0.05805708,0.13640099){\color[rgb]{0,0,0}\makebox(0,0)[b]{\smash{$u$}}}%
    \put(0.40667698,0.0755662){\color[rgb]{0,0,0}\makebox(0,0)[b]{\smash{$d_{vw}$}}}%
    \put(0,0){\includegraphics[width=\unitlength,page=2]{circuit_graph.pdf}}%
    \put(0.22135573,0.34648132){\color[rgb]{0,0,0}\makebox(0,0)[b]{\smash{comb. delay?}}}%
    \put(0,0){\includegraphics[width=\unitlength,page=3]{circuit_graph.pdf}}%
    \put(0.55545081,0.26256884){\color[rgb]{0,0,0}\makebox(0,0)[b]{\smash{seq. unit?}}}%
    \put(0,0){\includegraphics[width=\unitlength,page=4]{circuit_graph.pdf}}%
    \put(0.81020976,0.29187696){\color[rgb]{0,0,0}\makebox(0,0)[b]{\smash{anchor?}}}%
    \put(0.80435193,0.23485238){\color[rgb]{0,0,0}\makebox(0,0)[b]{\smash{$\lambda_{tz}$}}}%
    \put(0.23075348,0.29370577){\color[rgb]{0,0,0}\makebox(0,0)[b]{\smash{$\xi_{uv}$}}}%
    \put(0.4109505,0.01032405){\color[rgb]{0,0,0}\makebox(0,0)[b]{\smash{sizing?}}}%
    \put(0.51072335,0.11668065){\color[rgb]{0,0,0}\makebox(0,0)[b]{\smash{$w$}}}%
    \put(0.66035182,0.11668065){\color[rgb]{0,0,0}\makebox(0,0)[b]{\smash{$t$}}}%
    \put(0.9370643,0.11668065){\color[rgb]{0,0,0}\makebox(0,0)[b]{\smash{$z$}}}%
    \put(0,0){\includegraphics[width=\unitlength,page=5]{circuit_graph.pdf}}%
    \put(0.20768553,0.13640099){\color[rgb]{0,0,0}\makebox(0,0)[b]{\smash{$v$}}}%
  \end{picture}%
\endgroup%

\caption{Delay insertion model in \titlenym.}
\label{fig:circuit_graph}
}
\end{figure}

\hfill \break
\indent In \figname~\ref{fig:circuit_graph}, the delay at the circuit node can be
changed by sizing the delay of the logic gate, e.g., the XOR gate in \figname~\ref{fig:circuit_graph}. 
For the case that the required
gate delay exceeds the largest permissible value, a combinational delay unit is inserted at 
the corresponding input. For convenience, we assume the combinational delay unit inserted 
at the input is implemented with buffers. 
The relation between the arrival times
$u$ and $w$ is thus expressed as
\begin{align} \label{eq:no_ff_orig}
      s_{w} \ge s_{u}+\xi_{uv}*r^u+d_{vw}*r^u \\
      s^\prime_{w} \le s^\prime_{u}+\xi_{uv}* r^l+d_{vw}*r^l \label{eq:no_ff_orig1}
\end{align}
where $s_{u}$, $s^\prime_{u}$, $s_{w}$ and  $s^\prime_{w}$ 
are the latest and earliest arrival times of node $u$ and $w$ , respectively.
\reviewhl{$\xi_{uv}$ is the extra delay 
introduced by an inserted buffer and $d_{vw}$ is the pin-to-pin delay of the logic gate. }
If $\xi_{uv}$ is reduced to 0 
after optimization,
no buffer is required in the optimized circuit.
The $\le$ and $\ge$ relaxations of the relation between arrival times guarantee
that only the latest and the earliest arrival times from multiple inputs
are propagated further. 
$r^u$ and $r^l$ are two constants to reserve a guard band for process variations,
so that $r^u >1 $ and $r^l<1$.

Rising and falling pin-to-pin delays of logic gates, e.g., the XOR gate in \figname~\ref{fig:circuit_graph}, 
\reviewhl{might be different}. %
Using the rising and falling pin-to-pin delays of logic gates 
to separately establish the constraints of arrival times in 
(\ref{eq:no_ff_orig}) and (\ref{eq:no_ff_orig1}) incurs high computation complexity, and thus long execution time. 
For simplification, we evaluate $d_{vw}$ with  
the average of the rising and falling pin-to-pin delays of the logic gate. 
This simplification may lead to inaccuracy in 
evaluating arrival times of signals and thus affect the subsequent timing optimization negatively. To compensate 
this inaccuracy, we calibrate the arrival times of signals with commercial tools, which 
will be explained in Section~\ref{sec:first_inte}.

\subsubsection{Insertion of sequential delay units}\label{sec:seq}
\hfill \break
\indent Since arrival times through long and short paths reaching $w$ may have a
large difference, we may need to insert sequential delay units to delay 
the fast signal more than the slow signal. This can be implemented with the
sequential units shown in \figname~\ref{fig:difference}, where the gap between
the arrival times is reduced after passing a sequential delay unit, either a
flip-flop or a latch. To insert a sequential delay unit, three cases need to
be examined. \reviewhl{These cases have an ``either-or" relationship, which can be converted into 
equivalent linear forms \cite{gurobi}.}

\textbf{Case 1:} No sequential delay unit is inserted between $w$ and $t$ in
\figname~\ref{fig:circuit_graph}, so that
\begin{align}  \label{eq:ff_insertion_0}
  s_t\ge s_w\\
  s_t^\prime\le s^\prime_w.
\end{align}

\textbf{Case2:}  A flip-flop is inserted between $w$ and $t$. Assume the
flip-flop works at a rising clock edge. 
As shown in \figname~\ref{fig:difference}(b), a flip-flop only works properly 
 in a region $t_h$ after the rising clock edge and $t_{su}$ before the next
rising clock edge. 
Therefore, we need to bound  
the arrival times $s_w$ and $s^\prime_w$ into such a region by
\begin{align} \label{eq:ff_insertion_1}
      s_{w},s^\prime_{w} \ge N_{wt}*T+\phi_{wt}+t_{h}*r^u \\ 
      s_{w},s^\prime_{w} \le (N_{wt}+1)*T+\phi_{wt}-t_{su}*r^u
\label{eq:ff_insertion_2}
\end{align}
\reviewhl{where $N_{wt}$ is an integer variable whose value is determined by the solver. 
This variable 
represents that
the signal arrival time of $w$ 
can be within any clock cycle, 
the starting and ending time of which are $N_{wt}*T$ and $(N_{wt}+1)*T$, respectively. 
}  
$T$ is the given clock period. 
$\phi_{wt}$ is phase shift
of the clock signal. The available values of $\phi_{wt}$ can be set by designers.
If only one clock signal is available, $\phi_{wt}$ can be set to 0 and T/2 to
emulate flip-flops working at rising and falling clock edges. 

When the input arrival times fall into the valid region of a flip-flop as
constrained by (\ref{eq:ff_insertion_1})--(\ref{eq:ff_insertion_2}), 
the signal always starts to propagate from the next active clock edge, 
so that the constraints can be written as
\begin{align}
  s_t \ge (N_{wt}+1)T +\phi_{wt}+t_{cq}*r^u\\ 
  s_t^\prime\le (N_{wt}+1)T +\phi_{wt}+t_{cq}*r^l.
\end{align}

\textbf{Case3:} A level-sensitive latch is inserted between $w$ and $t$.
To be consistent with the active region of flip-flops, we assume that the
latches are transparent when the clock signal is equal to 0. We can then
bound the arrival times at $w$ the same as 
(\ref{eq:ff_insertion_1})--(\ref{eq:ff_insertion_2}).

As illustrated in \figname~\ref{fig:difference}(c), the latch is
non-transparent in the first part of the region and transparent in the 
second region. Accordingly, the latest time a signal leaves the latch can be expressed as
\begin{align} \label{eq:arriv_latch_orig_1}
s_t\ge N_{wt}*T+\phi_{wt}+D*T+t_{cq}*r^u\\ 
s_t\ge s_w+t_{dq}*r^u
\label{eq:arriv_latch_orig_2}
\end{align}
where (\ref{eq:arriv_latch_orig_1}) corresponds to the case that the latch is
non-transparent, so that the signal leaves the latch at the moment the clock
switches to 1. 
$D$ is the duty cycle of the clock signal with $0<D<1$.
(\ref{eq:arriv_latch_orig_2}) corresponds to the case that the latch is
transparent, so that only the delay of the latch is added to $s_w$. 
$t_{dq}$ is the
data-to-q delay of the latch.

The earliest time a signal leaves the latch is, however, imposed by a constraint in the
less-than-max form as in \cite{SAMO90},
\begin{equation}\label{eq:latch_min_max}
  s^\prime_t\le \max\{N_{wt}*T+\phi_{wt}+D*T+t_{cq}*r^l,\;\; s^\prime_w+t_{dq}*r^l\}
\end{equation}
which cannot be linearized easily. In the \titlenym\ framework, the purpose of
introducing the sequential delay unit is to delay the short path as much as
possible. This effect happens when a signal arrives at a non-transparent latch. 
Therefore, we impose the arrival times of fast signals to be positioned 
in the non-transparent region, expressed as 
\begin{equation}\label{eq:latch_new_region}
N_{wt}*T+\phi_{wt}+t_{h}*r^u  \le s^\prime_{w} \le N_{wt}*T+\phi_{wt}+D*T
\end{equation}
while relaxing (\ref{eq:latch_min_max}) as
\begin{equation}\label{eq:latch_min_max_bound}
  s^\prime_t\le N_{wt}*T+\phi_{wt}+D*T+t_{cq}*r^l.
\end{equation}

When inserting the sequential delay unit, each of the three cases above can happen
in the optimized circuit. We use an integer variable to represent the
selection and let the solver determine which case happens during the optimization.

\subsubsection{Reference shifting with respect to anchor points}\label{sec:reference}
\hfill \break
\indent The arrival times in the model need to be converted each time when an anchor point is
passed. The constant $\lambda_{tz}$ represents the number of flip-flops at such a point in the original
circuit. In \figname~\ref{fig:circuit_graph}, the arrival time at $z$ is
shifted as
\begin{equation}
  s_z=s_t-\lambda_{tz}T.
\end{equation}

\subsubsection{Wave non-interference condition} \label{sec:wave}
\hfill \break
\indent \reviewhl{Since we allow multiple waves to propagate along a combinational path, we need to
guarantee that the signal of the next wave starting from a boundary flip-flop  never catches the
signal of the previous wave starting from the same flip-flop %
\cite{Burleson1998}.} This constraint should be imposed to every node in the circuit.
For example, the constraint for node
$u$ is written as 
\begin{align} \label{eq:stable_orig}
       s_{u}+t_{stable}\le s^\prime_{u}+T
\end{align}  
where
$t_{stable}$ is the minimum gap 
between two consecutive signals.

\subsubsection{Overall formulation}
\hfill \break
\indent The introduction of the relative timing references, or the anchor points,
in Section~\ref{sec:relative_timing} guarantees that the number of clock cycles
along any path does not change after optimization. %
With the timing constraints
(\ref{eq:simple_setup})-(\ref{eq:simple_hold}) at boundary flip-flops,
the correct function of the optimized circuit is always maintained, without
requiring any change in other function blocks.

\reviewhl{The constraints (\ref{eq:simple_setup})--(\ref{eq:stable_orig}) excluding
(\ref{eq:latch_min_max}) need to be established at each node in the circuit
after flip-flops are removed to 
guarantee the correct timing interface to flip-flops at the boundary of the optimized circuit parts. 
If this timing interface %
is correct, 
the functionality of the circuit is maintained \cite{Burleson1998,Joy1993, Yehuda2021}}. The appearance of the combinational and
sequential delay units needs to be 
determined by the solver. The delays of logic gates should also be sized.  The
objective of the optimization is to find a solution to make the circuit work
at a given clock period $T$, while reducing the area cost.  
Taking all these
factors into account, the straightforward ILP formulation may become
insolvable. In practice, however, this technique only needs to be applied to
isolated circuit parts containing critical paths. In addition, we introduce
heuristic techniques to overcome this scalability problem, 
as explained in the following section.

\section{Timing Optimization with Relaxed \titlenym\ Timing Model}\label{sec:impl}

\begin{figure}[t]
{
\figurefontsize
\centering
\begingroup%
  \makeatletter%
  \providecommand\color[2][]{%
    \errmessage{(Inkscape) Color is used for the text in Inkscape, but the package 'color.sty' is not loaded}%
    \renewcommand\color[2][]{}%
  }%
  \providecommand\transparent[1]{%
    \errmessage{(Inkscape) Transparency is used (non-zero) for the text in Inkscape, but the package 'transparent.sty' is not loaded}%
    \renewcommand\transparent[1]{}%
  }%
  \providecommand\rotatebox[2]{#2}%
  \ifx\svgwidth\undefined%
    \setlength{\unitlength}{200.74397009bp}%
    \ifx\svgscale\undefined%
      \relax%
    \else%
      \setlength{\unitlength}{\unitlength * \real{\svgscale}}%
    \fi%
  \else%
    \setlength{\unitlength}{\svgwidth}%
  \fi%
  \global\let\svgwidth\undefined%
  \global\let\svgscale\undefined%
  \makeatother%
  \begin{picture}(1,1.14086894)%
    \put(0,0){\includegraphics[width=\unitlength,page=1]{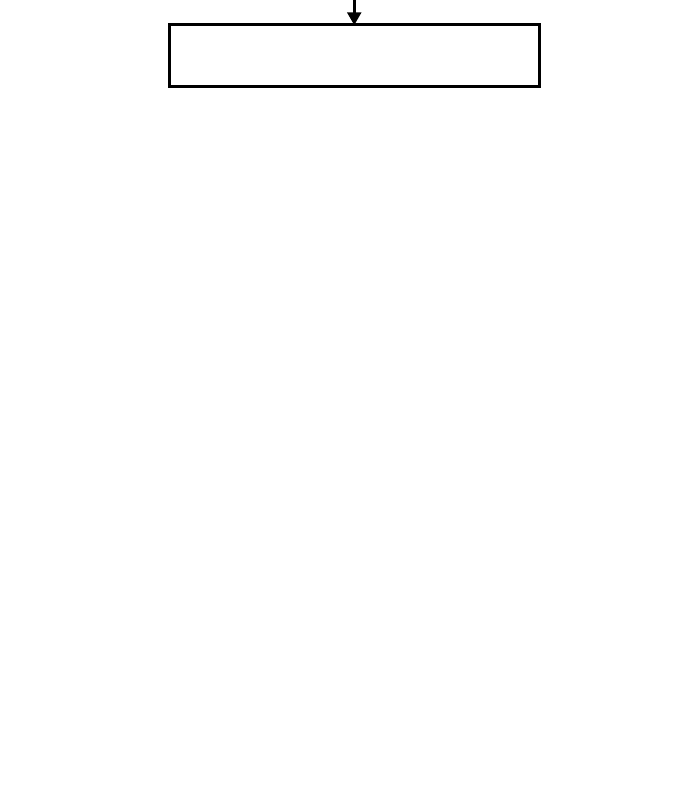}}%
    \put(0.50779046,1.07085731){\color[rgb]{0,0,0}\makebox(0,0)[b]{\smash{Emulation of sequential delays}}}%
    \put(0.50852987,1.03368304){\color[rgb]{0,0,0}\makebox(0,0)[b]{\smash{with $\delta_{wt}^\prime-\delta_{wt}$ }}}%
    \put(0,0){\includegraphics[width=\unitlength,page=2]{flow.pdf}}%
    \put(0.51243066,0.78685723){\color[rgb]{0,0,0}\makebox(0,0)[b]{\smash{All $\delta_{wt}^\prime-\delta_{wt}=0$}}}%
    \put(0.54159104,0.71135012){\color[rgb]{0,0,0}\makebox(0,0)[b]{\smash{No}}}%
    \put(0,0){\includegraphics[width=\unitlength,page=3]{flow.pdf}}%
    \put(0.27642447,0.32976109){\color[rgb]{0,0,0}\makebox(0,0)[b]{\smash{No}}}%
    \put(0.5456429,0.2348036){\color[rgb]{0,0,0}\makebox(0,0)[b]{\smash{Yes}}}%
    \put(0,0){\includegraphics[width=\unitlength,page=4]{flow.pdf}}%
    \put(0.50794236,0.02430683){\color[rgb]{0,0,0}\makebox(0,0)[b]{\smash{Optimized circuit}}}%
    \put(0.50779027,0.92530964){\color[rgb]{0,0,0}\makebox(0,0)[b]{\smash{Model approximation with clock/data-to-q delays}}}%
    \put(0.50735229,0.1740462){\color[rgb]{0,0,0}\makebox(0,0)[b]{\smash{Buffer replacement using sequential delay units }}}%
    \put(0.50845428,0.13325783){\color[rgb]{0,0,0}\makebox(0,0)[b]{\smash{and delay discretization}}}%
    \put(0,0){\includegraphics[width=\unitlength,page=5]{flow.pdf}}%
    \put(0.5071787,0.65535822){\color[rgb]{0,0,0}\makebox(0,0)[b]{\smash{Decrease the lower bound of }}}%
    \put(0.50748679,0.61720912){\color[rgb]{0,0,0}\makebox(0,0)[b]{\smash{$\delta_{wt}^\prime-\delta_{wt}$ }}}%
    \put(0,0){\includegraphics[width=\unitlength,page=6]{flow.pdf}}%
    \put(0.74324577,0.80565154){\color[rgb]{0,0,0}\makebox(0,0)[b]{\smash{Yes}}}%
    \put(0.50779045,0.47927026){\color[rgb]{0,0,0}\makebox(0,0)[b]{\smash{Model legalization using accurate delay models}}}%
    \put(0.50843265,0.44002889){\color[rgb]{0,0,0}\makebox(0,0)[b]{\smash{and update $S_d$}}}%
    \put(0,0){\includegraphics[width=\unitlength,page=7]{flow.pdf}}%
    \put(0.51306139,0.30944336){\color[rgb]{0,0,0}\makebox(0,0)[b]{\smash{All $\delta_{wt}^\prime-\delta_{wt}=0$}}}%
    \put(0,0){\includegraphics[width=\unitlength,page=8]{flow.pdf}}%
  \end{picture}%
\endgroup%

\caption{The proposed timing optimization with relaxed \titlenym\ flow.}
\label{fig:flow}
}
\end{figure}

In applying the \titlenym\ timing model above, we introduce a framework to identify the
locations of delay unit with iterative relaxation of \titlenym.  %
The flow of this framework is shown in \figname~\ref{fig:flow}, which will be 
explained in this section.  
The basic strategy is to remove flip-flops along critical paths to 
eliminate the inherent clock-to-q delays and setup time. 
Thereafter, fast signals of short paths will be blocked to guarantee the correct functionality by inserting the 
minimum number of sequential delay units and buffers. 
To reduce area overhead, buffers are replaced with sequential delay units.   

\subsection{Emulation of Sequential Delay Units}\label{sec:first_step}

After we remove all
the flip-flops from the original circuit, the short paths may have extremely
small delays. The gap between these delays and those of long paths is very large. 
It cannot be 
reduced with combinational delay units since they introduce the same delays to the fast and slow signals as shown in \figname~\ref{fig:difference}(a). 
Instead, only sequential delay units are
able to reduce this gap so that the fast and slow signals still arrive at
boundary flip-flops within the same clock cycle as those of the original circuit.

In the first step of the framework, we identify the locations at which
sequential delay units are indispensable. Without these units, the fast and
slow signals, such as those within feedback loops, may not be aligned properly into the correct clock cycles, even
though unlimited combinational delay units can be inserted into the circuit. 
In practice, however, it is not easy to identify the exact locations of these units using the exact and complete model in (\ref{eq:ff_insertion_0})--(\ref{eq:latch_min_max_bound}) directly.
To solve this problem, 
we relax the timing constraints (\ref{eq:ff_insertion_0})--(\ref{eq:latch_min_max_bound}) 
instead of modeling them directly. 
\reviewhl{From \figname~\ref{fig:difference}, 
we can see that a buffer slows down fast and slow signals with the same delay. On the contrary, 
a sequential delay unit slows down fast and slow signals with different delays. 
For example, no matter when fast and slow signals arrive at the input of a flip-flop, they leave the output of the flip-flop at the same time. 
Accordingly, the flip-flop slows down the fast signal with a larger delay than the slow signal. 
Taking advantage of this characteristic, 
we use two non-negative variables $\delta_{wt}$ and $\delta^\prime_{wt}$ to emulate the delay effects of three delay units 
as described above.  %
In case $\delta_{wt} = \delta^\prime_{wt}$, a buffer is inserted. In case $\delta_{wt} < \delta^\prime_{wt}$, a 
sequential delay unit is inserted. 
}
When signals travel
from $u$ to $z$ in \figname~\ref{fig:circuit_graph}, the relation of arrival times  
from nodes $u$ to $z$ can be written as
\begin{align} 
  s_{z} \ge s_{u} +\xi_{uv}*r^u +d_{vw}*r^u +\delta_{wt}-\lambda_{tz}T \label{eq:ff_orig1} \\ 
  s^\prime_{z} \le s^\prime_{u} +\xi_{uv}*r^l +d_{vw}*r^l+
  \delta^\prime_{wt}-\lambda_{tz}T \label{eq:ff_orig2}  \\
 \reviewhl{0 \le  \delta_{wt} \le \delta_{wt}^\prime}\label{eq:delta_relation}\\
   s^\prime_{u}+\delta_{wt}^\prime \le s_{u}+\delta_{wt} \label{eq:delta_arr_relation}
\end{align}
where the variables $\delta_{wt}$ and $\delta_{wt}^\prime$ emulate delays introduced by sequential delay units. 
(\ref{eq:delta_relation}) specifies that
the fast signal should be padded with more delays than the slow signal.  
\reviewhl{ %
The purpose of (\ref{eq:delta_arr_relation}) 
is to guarantee that a sequential delay unit is required to break a feedback loop.
}

The optimization problem to find the potential locations of sequential delay
units is thus written as
\begin{align} \label{eq:obj_orig1}
\text{minimize}  \quad& \alpha\sum_{G}(\delta_{wt}^\prime-\delta_{wt})+
\beta\sum_{G}(\delta_{wt}^\prime+\xi_{uv})-\gamma\sum_{G}{d_{vw}}\\
\text{subject to}  \quad&
\text{(\ref{eq:stable_orig})--(\ref{eq:delta_arr_relation}) for each gate in $G$}\\
\phantom{subject to}  \quad&
\text{constr. (\ref{eq:simple_setup})--(\ref{eq:simple_hold})
for each boundary flip-flop}\label{eq:opt_1_bnd}
\end{align}
\reviewhl{where $G$ is the set of all logic gates in the original circuit. 
This optimization problem also maximizes the overall delays of logic gates in
the circuit with the last term in (\ref{eq:obj_orig1}) since larger delays indicate smaller area. %
Since the area of a flip-flop is
about 6 times of the area of a buffer and the average area of combinational gates, 
$\alpha$, $\beta$ and $\gamma$ are set to 100,10,10, to specify the balance
between the area of sequential delay units, inserted buffers and logic gates. 
With this setting, we provide the solver a clear tendency to minimize the area of sequential delay units, buffers and combinational gates with different
priorities. 
In the second term, $\delta_{wt}^\prime$ also represents the existence of buffers in case $\delta_{wt}^\prime = \delta_{wt}$.}
Solving the optimization problem above identifies nodes with unequal padding delays
$\delta_{wt}$ and $\delta_{wt}^\prime$, indicating potential locations of
sequential delay units, as a set $S$. These delays may still violate the exact constraints in (\ref{eq:ff_insertion_0})--(\ref{eq:latch_min_max_bound}), 
so that they need to be refined further.

\subsection{Modeling with Clock/Data-to-Q Delays of Sequential Delay Units} 
\label{sec:second_step}

The optimization problem (\ref{eq:obj_orig1})--(\ref{eq:opt_1_bnd}) does not 
consider the inherent clock-to-q delays of flip-flops and data-to-q delays of
latches. Since these delays are introduced only at locations where
sequential delay units are inserted, they need to be modeled for all the locations 
$S$ returned by the previous step. We introduce a binary variable $x_{wt}$ to
represent whether a sequential delay unit appears at a location from $S$, and
revise the constraints (\ref{eq:ff_orig1})--(\ref{eq:ff_orig2}) as
\begin{align} 
  & s_{z} \ge s_{u} +\xi_{uv}*r^u +d_{vw}*r^u +x_{wt}\delta_{wt}+
  x_{wt}t_{cd\to q}*r^u-\lambda_{tz}T \label{eq:ff_delay_orig1} \\ 
  & s^\prime_{z} \le s^\prime_{u} +\xi_{uv}*r^l +d_{vw}*r^l
  +x_{wt}\delta^\prime_{wt}+x_{wt}t_{cd\to q}*r^l-\lambda_{tz}T
  \label{eq:ff_delay_orig2}  
\end{align}
where $t_{cd\to q}$ represents
clock-to-q delay or data-to-q delay, which can be evaluated according to the output 
load of the sequential components, and 
the delays $\delta_{wt}$, $\delta^\prime_{wt}$ and $t_{cd\to q}$ are only valid 
when $x_{wt}$ is equal to 1. The inclusion of the binary variables $x_{wt}$ is very
computation-intensive, so that they can only be dealt with after the potential
locations of sequential delay units are reduced to $S$ by solving
(\ref{eq:obj_orig1})--(\ref{eq:opt_1_bnd}). 
Since $x_{wt}$ is a binary variable, the multiplications $x_{wt}\delta_{wt}$ and 
$x_{wt}\delta^\prime_{wt}$ can be converted into equivalent linear forms so
that the overall formulation is still an ILP problem.

Considering the inherent delays of sequential delay units, their locations can
be refined further by solving the optimization problem as
\begin{align} \label{eq:obj_delay_orig1}
\text{minimize}  \quad& \alpha\sum_{G/S}(\delta_{wt}^\prime-\delta_{wt})+
\beta\sum_{G/S}(\delta_{wt}^\prime+\xi_{uv})
-\gamma\sum_{G}{d_{vw}}\\
\text{subject to}\quad  &%
\text{(\ref{eq:stable_orig})--(\ref{eq:delta_arr_relation}) for each gate in \reviewhl{$G\backslash S$}}\\
\phantom{subject to} & %
\text{(\ref{eq:stable_orig}), (\ref{eq:delta_relation})--(\ref{eq:delta_arr_relation}) 
for each gate in $S$}\\
\phantom{subject to} & %
\text{(\ref{eq:ff_delay_orig1})--(\ref{eq:ff_delay_orig2})
for each gate in $S$
}\\
\phantom{subject to} & %
\text{constr. (\ref{eq:simple_setup})--(\ref{eq:simple_hold})
for each boundary flip-flop}\label{eq:opt_2_bnd}\\
\phantom{subject to} &
\reviewhl{\delta_{wt}^\prime-\delta_{wt} \ge d_{th} \text{ if } x_{wt}=1 \label{eq:obj_delay_orig1_bnd}}
\end{align}
\reviewhl{where $d_{th}$ is a predefined 
delay bound. %
The condition in (32) can be converted into equivalent linear forms according to \cite{gurobi}. }

\reviewhl{To identify the necessary locations of sequential delay units, we execute the 
optimization (\ref{eq:obj_delay_orig1})--(\ref{eq:obj_delay_orig1_bnd}) iteratively until 
no different $\delta_{wt}$ and $\delta_{wt}^\prime$ exist anymore. %
In each iteration, we 
lower the delay bound $d_{th}$ linearly. %
A large bound of $\delta_{wt}^\prime-\delta_{wt}$ in the early iterations allows
the solver to quickly determine the important locations for inserting sequential delay units.
The refined
locations of sequential delay units from this step are returned as a set
$S_d$, which is more accurate than $S$.}

\subsection{Model Legalization for Timing of Sequential Delay Units} \label{sec:third_step}

In this step, the complete model described in Section~\ref{sec:mode_delay_unit} is
applied to the locations in $S_d$ to generate sequential delay units that are
really required in the circuit. The optimization problem is described as 
\begin{align} \label{eq:obj_model_orig1}
\text{minimize}  \quad& \alpha\sum_{G/S_d}(\delta_{wt}^\prime-\delta_{wt})+
\beta\sum_{G/S_d}(\delta_{wt}^\prime+\xi_{uv})
-\gamma\sum_{G}{d_{vw}}\\
\text{subject to}  \quad&
\text{(\ref{eq:stable_orig})--(\ref{eq:delta_arr_relation}) for each gate in \reviewhl{$G\backslash S_{d}$}}\\
\phantom{subject to}  \quad&
\text{(\ref{eq:ff_insertion_0})--(\ref{eq:arriv_latch_orig_2}) 
for each gate in $S_d$}\\
\phantom{subject to}  \quad&
\text{(\ref{eq:latch_new_region})--(\ref{eq:stable_orig})
for each gate in $S_d$ }\\
\phantom{subject to}  \quad&
\text{constr. (\ref{eq:simple_setup})--(\ref{eq:simple_hold})
for each boundary flip-flop.}\label{eq:opt_3_bnd}
\end{align}

After solving the optimization above, there might still be different
$\delta_{wt}$ and $\delta^\prime_{wt}$ in $G\backslash S_d$, \reviewhl{because the timing legalization
of sequential delay units with the complete model (\ref{eq:simple_setup})--(\ref{eq:stable_orig}) excluding
(\ref{eq:latch_min_max}) in Section~\ref{sec:mode_delay_unit} 
may invalidate some locations in $S_d$ so that $S_d$ reduces in size.} The complete %
model is applied to these locations iteratively, 
until no different $\delta_{wt}$ and $\delta^\prime_{wt}$ exists, indicating
the remaining timing synchronization can be achieved with buffers and gate
sizing directly.

\subsection{Buffer Replacement with Sequential Units} \label{sec:four_step}

After solving (\ref{eq:obj_model_orig1})--(\ref{eq:opt_3_bnd}), delays $d_{vw}$ of logic gates are 
set to the nearest discrete delay values defined in the library.
Buffer delays $\xi_{uv}$ are also determined. 
If $\xi_{uv}$ is large, several buffers are needed for its implementation.
 As shown in \figname~\ref{fig:difference},
sequential delay units can introduce a very large delay. For example, a flip-flop
can introduce a delay as large as $T+t_{cq}-t_h$, if the incoming signal
arrives at the flip-flop right after a clock edge.  According to this
observation, we iteratively replace buffers with large delays by sequential delay units
to reduce area. 
In each iteration, the accurate sequential model
(\ref{eq:ff_insertion_0})--(\ref{eq:arriv_latch_orig_2}) and
(\ref{eq:latch_new_region})--(\ref{eq:stable_orig}) is applied to guarantee these new sequential delay units are
valid. The iteration stops when no buffer can be replaced by sequential units. 
Buffers that cannot be replaced by sequential delay 
units are implemented directly in the optimized circuit.

\nopagebreak

\section{Circuit Fine-Tuning with \titlenym\ in Commercial Tools}\label{sec:integration}

The timing optimization with relaxed \titlenym\ described in Section~\ref{sec:impl} 
can generate the optimized circuits where timing performance is improved and area cost is reduced. 
However, this method might be inconsistent when it is interfaced with commercial tools, e.g., Design Compiler from Synopsys. 
For example, 
the average pin-to-pin delays are used in evaluating the arrival times
of signals for the sake of execution efficiency in the relaxed \titlenym. 
Therefore, %
we fine-tune the optimized circuits with Design Compiler. 

To fine-tune the optimized circuits with Design Compiler, there are 
several challenges. First, 
the arrival times of signals should be evaluated 
accurately to make the optimization effective. 
Second, after the optimization, 
the removal locations of flip-flops in the optimized circuits should be extracted 
to establish 
the wave-pipelining timing constraints, with which  
Design Compiler can fine-tune the circuits. 
Third, 
buffers should be adjusted along short paths to satisfy the wave-pipelining timing constraints.

To overcome the challenges described above, %
we first adjust the iterative relaxation method where iterations of delay unit insertion are executed, as described in Section~\ref{sec:impl},  
to compensate the inaccuracy in timing optimization. Afterwards, 
we extract the removal locations of flip-flops with respect to the circuits under optimization  
and establish the corresponding wave-pipelining constraints compatible with Design Compiler. 
These timing constraints are then incorporated into the commercial tools to 
optimize the circuits. 
After this optimization, the delays of short paths might be still too small to meet the wave-pipelining constraints, 
so that 
buffers are inserted to pad their delays until timing constraints are met. These steps 
can be automatically executed without manual intervention.

\subsection{Adjusting Iterative Relaxation for Fine-Tuning of \titlenym}\label{sec:first_inte}

During the iterative relaxation %
described 
in Section~\ref{sec:impl}, 
the average pin-to-pin delays are used in evaluating the arrival times of signals for the sake of execution efficiency. 
However, this simplification might incur inaccuracy in evaluating the arrival times of signals and thus path delays. 
Therefore, 
critical paths 
might be considered as non-critical and thus  
cannot be optimized effectively.
To 
solve this problem, we adjust the iterative relaxation in Section~\ref{sec:impl} to compensate this 
evaluation inaccuracy in timing optimization. 
In this adjustment,  
we first run timing slack analysis for all the flip-flops of a circuit with Design Compiler 
before executing the iterative relaxation framework.
Afterwards, the latest and earliest arrival times of signals at the inputs of flip-flops evaluated with Design Compiler 
are used to calibrate those evaluated with the average pin-to-pin delays. 
Specifically, we record the gaps between arrival times evaluated with Design Compiler and with average pin-to-pin delays. 
During the optimization with the iterative relaxation, 
the arrival times at the inputs of flip-flops evaluated with average pin-to-pin delays are calibrated by adding
the corresponding gaps. %
With this  calibration, the critical paths can be identified and optimized effectively. 

During the timing optimization with the iterative relaxation framework, 
gate sizing plays an important role of increasing/decreasing delays for short/long paths to meet the timing constraints.
Since Design Compiler has the advantages of high accuracy and execution efficiency 
in sizing gates while minimizing the area overhead, 
we shift the task of gate sizing to Design Compiler 
in the adjusted iterative relaxation framework. 
Specifically, 
we first adopt the adjusted iterative relaxation without gate sizing %
to insert necessary sequential delay units to block fast signals along short paths in a circuit.   
Afterwards, we use Design Compiler to size gates to increase/decrease delays of short/long paths. 
If the delays of short paths are still too small after gate sizing, 
buffers are inserted to pad their delays, as explained later. %

\subsection{Extracting Removal Locations of Flip-flops }\label{sec:second_inte}
\begin{figure}[t]
{
\figurefontsize
\centering
\begingroup%
  \makeatletter%
  \providecommand\color[2][]{%
    \errmessage{(Inkscape) Color is used for the text in Inkscape, but the package 'color.sty' is not loaded}%
    \renewcommand\color[2][]{}%
  }%
  \providecommand\transparent[1]{%
    \errmessage{(Inkscape) Transparency is used (non-zero) for the text in Inkscape, but the package 'transparent.sty' is not loaded}%
    \renewcommand\transparent[1]{}%
  }%
  \providecommand\rotatebox[2]{#2}%
  \ifx\svgwidth\undefined%
    \setlength{\unitlength}{195.93490167bp}%
    \ifx\svgscale\undefined%
      \relax%
    \else%
      \setlength{\unitlength}{\unitlength * \real{\svgscale}}%
    \fi%
  \else%
    \setlength{\unitlength}{\svgwidth}%
  \fi%
  \global\let\svgwidth\undefined%
  \global\let\svgscale\undefined%
  \makeatother%
  \begin{picture}(1,0.98673501)%
    \put(0.49165894,0.51619106){\color[rgb]{0,0,0}\makebox(0,0)[b]{\smash{(a)}}}%
    \put(0.49165894,0.00640566){\color[rgb]{0,0,0}\makebox(0,0)[b]{\smash{(b)}}}%
    \put(0,0){\includegraphics[width=\unitlength,page=1]{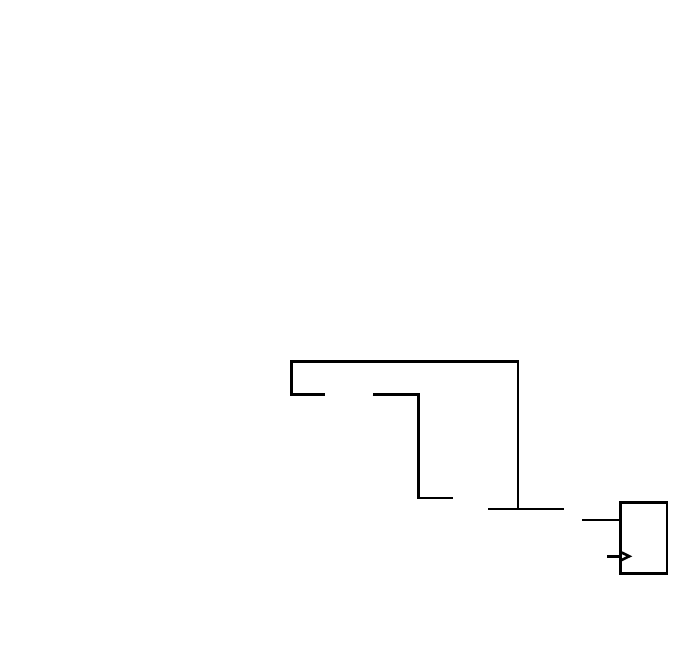}}%
    \put(0.94534681,0.18403055){\color[rgb]{0,0,0}\makebox(0,0)[b]{\smash{F4}}}%
    \put(0,0){\includegraphics[width=\unitlength,page=2]{remove_ff1.pdf}}%
    \put(0.51230831,0.36590165){\color[rgb]{0,0,0}\makebox(0,0)[b]{\smash{F6}}}%
    \put(0,0){\includegraphics[width=\unitlength,page=3]{remove_ff1.pdf}}%
    \put(0.2402615,0.09235221){\color[rgb]{1,0,0}\makebox(0,0)[lt]{\begin{minipage}{0.1556333\unitlength}\centering anchor point\end{minipage}}}%
    \put(0,0){\includegraphics[width=\unitlength,page=4]{remove_ff1.pdf}}%
    \put(0.05679885,0.25123443){\color[rgb]{0,0,0}\makebox(0,0)[b]{\smash{F2}}}%
    \put(0,0){\includegraphics[width=\unitlength,page=5]{remove_ff1.pdf}}%
    \put(0.05706863,0.0967252){\color[rgb]{0,0,0}\makebox(0,0)[b]{\smash{F1}}}%
    \put(0,0){\includegraphics[width=\unitlength,page=6]{remove_ff1.pdf}}%
    \put(0.05672692,0.39110436){\color[rgb]{0,0,0}\makebox(0,0)[b]{\smash{F3}}}%
    \put(0,0){\includegraphics[width=\unitlength,page=7]{remove_ff1.pdf}}%
    \put(0.42373242,0.21583776){\color[rgb]{1,0,0}\makebox(0,0)[lt]{\begin{minipage}{0.1556333\unitlength}\centering anchor  point\end{minipage}}}%
    \put(0,0){\includegraphics[width=\unitlength,page=8]{remove_ff1.pdf}}%
    \put(0.60161419,0.13397886){\color[rgb]{1,0,0}\makebox(0,0)[lt]{\begin{minipage}{0.1556333\unitlength}\centering anchor point\end{minipage}}}%
    \put(0,0){\includegraphics[width=\unitlength,page=9]{remove_ff1.pdf}}%
    \put(0.68945715,0.72061794){\color[rgb]{0,0,0}\makebox(0,0)[b]{\smash{F6}}}%
    \put(0,0){\includegraphics[width=\unitlength,page=10]{remove_ff1.pdf}}%
    \put(0.31498907,0.60521648){\color[rgb]{0,0,0}\makebox(0,0)[b]{\smash{F5}}}%
    \put(0,0){\includegraphics[width=\unitlength,page=11]{remove_ff1.pdf}}%
    \put(0.05367649,0.91581195){\color[rgb]{0,0,0}\makebox(0,0)[b]{\smash{F3}}}%
    \put(0,0){\includegraphics[width=\unitlength,page=12]{remove_ff1.pdf}}%
    \put(0.94544686,0.70650413){\color[rgb]{0,0,0}\makebox(0,0)[b]{\smash{F4}}}%
    \put(0,0){\includegraphics[width=\unitlength,page=13]{remove_ff1.pdf}}%
    \put(0.05645709,0.77639545){\color[rgb]{0,0,0}\makebox(0,0)[b]{\smash{F2}}}%
    \put(0,0){\includegraphics[width=\unitlength,page=14]{remove_ff1.pdf}}%
    \put(0.05672682,0.62188636){\color[rgb]{0,0,0}\makebox(0,0)[b]{\smash{F1}}}%
    \put(0.05745171,0.54216166){\color[rgb]{0,0,0}\makebox(0,0)[b]{\smash{FF}}}%
    \put(0,0){\includegraphics[width=\unitlength,page=15]{remove_ff1.pdf}}%
    \put(0.05745171,0.01831475){\color[rgb]{0,0,0}\makebox(0,0)[b]{\smash{FF}}}%
  \end{picture}%
\endgroup%

\caption{The circuit under optimization and the optimized circuit after the adjusted iterative relaxation. (a) The circuit under optimization where $g_i$/ZN represent the output pin of $g_i$ and $g_i$/A, $g_i$/A1 and $g_i$/A2 are the input pins of $g_i$. (b) the optimized circuit where flip-flops are removed. }
\label{fig:remove_ff}
}
\end{figure}

After the adjusted iterative relaxation is executed, 
flip-flops along critical paths are removed and 
fast signals are blocked with sequential delay units in the optimized circuit. 
Therefore, 
the sequential components in the optimized circuit 
are reallocated with respect to the circuit under optimization. 
\figname~\ref{fig:remove_ff} 
illustrates a circuit under optimization  %
 and the optimized circuit where red dots represent anchor points, also the removal locations 
of flip-flops. 
By comparing the circuit under optimization in \figname~\ref{fig:remove_ff}(a) and the optimized  
circuit in \figname~\ref{fig:remove_ff}(b), we can see that  
F6 in the circuit under optimization  
is moved leftwards with retiming to block signals along a loop, 
while  %
the retimed flip-flops 
between $g_4$ and $g_5$, $g_2$ and $g_5$ and F5 are removed to allow slow signals to propagate through. 

The removal of flip-flops %
creates wave-pipelining paths where several logic waves propagate 
simultaneously in the optimized circuit. 
For example, in \figname~\ref{fig:remove_ff}(b), 
the wave-pipelining path from F2 to F4 has two logic waves propagating along it since this path traverses through one 
removal location along it and 
the wave-pipelining path from F1 to F4 has three logic waves 
propagating along it since it traverses through two removal locations along it. 
To guarantee the correct functionality of the optimized circuit, 
the delays of such paths should satisfy the corresponding wave-pipelining timing constraints. 
For example, %
the largest delay of the path between F1 and F4 should be smaller than $3\cdot T-t_{su}$, and
the smallest delay should be larger than $2\cdot T+t_{h}$,
where $T$ is the target clock period, $t_{su}$ and $t_{h}$ are the setup time and hold time of F4, respectively. 
To optimize such paths to satisfy the wave-pipelining constraints, %
we should 
establish the wave-pipelining constraints compatible with Design Compiler and then 
incorporate such constraints into the optimization flow of Design Compiler. 
To achieve this goal, 
we first extract the removal locations of flip-flops with respect to the circuit under optimization. %

According to \figname~\ref{fig:remove_ff}, 
the removal locations of flip-flops in the optimized circuit can be extracted 
by applying retiming together with removing flip-flops on the circuit under optimization. 
To apply this technique on the circuit under optimization, %
we use $g \in G$ to represent a combinational gate and $e_{g_i,g_j} \in E$ between gates $g_i$ and $g_j$ 
to represent the net connecting the output of the combinational gate $g_i$ and an
input of another combinational gate $g_j$. 
$e_{g_i,g_j}$ %
has a constant weight $w(e_{g_i,g_j})$ to represent the number of flip-flops along
the connection in the circuit under optimization.
Each combinational gate has a retiming variable $r(g)$, which defines how many flip-flops are moved from the
output of a gate to its inputs. After retiming, the number of flip-flops on a net between gates $g_i$ and $g_j$
is written as $w_r(e_{g_i,g_j})=w(e_{g_i,g_j})+r(g_j)-r(g_i)$. 
The number of flip-flops along $e_{g_i,g_j}$ in the optimized circuit 
is denoted as $w^\prime(e_{g_i,g_j})$, which can be different from $w_r(e_{g_i,g_j})$ due to the removal of flip-flops.  
To determine how many flip-flops along $e_{g_i,g_j}$ are removed in the optimized circuit, 
a variable $y_{e_{g_i,g_j}}$ is assigned for $e_{g_i,g_j}$. 
With this setting, two cases for a net between gates $g_i$ and $g_j$ should be examined.

\textbf{Case 1:} If the net from gate $g_i$ to gate $g_j$ has the retimed weight equal to the weight in the optimized circuit, namely 
$w_r(e_{g_i,g_j})=w(e_{g_i,g_j})+r(g_j)-r(g_i)= w^\prime(e_{g_i,g_j})$, 
there is no removal of flip-flops along this net, so that $y_{e_{g_i,g_j}}=0$. For example, 
$y_{e_{g_5,g_6}}=0$ in case of $r(g5)=1$. 

\textbf{Case 2:} If the net from gate $g_i$ to gate $g_j$ has the retimed weight larger than that in the optimized circuit, namely
$w_r(e_{g_i,g_j})=w(e_{g_i,g_j})+r(g_j)-r(g_i)\ge w^\prime(e_{g_i,g_j})+1$, 
flip-flops are removed along this net, and the number of removed flip-flops is $y_{e_{g_i,g_j}}=w_r(e_{g_i,g_j})-w^\prime(e_{g_i,g_j})$. 
For example, 
$y_{e_{g_4,g_5}}=1$, $y_{e_{g_2,g_5}}=1$ and $y_{e_{g_1,g_2}}=1$ .

When establishing the relation between the circuit under optimization and the optimized circuit, 
each of the cases above can happen. 
We use the constraints in the two cases described above 
to establish an ILP formulation %
and let the solver determine which case actually happens during %
the adjusted iterative relaxation. After that, 
we can obtain the removal locations of flip-flops 
along net connections where $y_{e_{g_i,g_j}} \ge 1$ in the optimized circuit. 

\subsection{Generating Wave-pipelining Timing Constraints for Integration of \titlenym}\label{sec:third_inte}

After the removal locations of flip-flops are extracted with the method in Section~\ref{sec:second_inte}, 
we can use them to automatically establish the wave-pipelining constraints compatible with Design Compiler. 
In Design Compiler, 
wave-pipelining timing constraints can be set with the following commands 
\begin{nospaceflalign}
&\text{\equationfontsize $set\_max\_delay$ $d$ -$from$ $p_1$ -$through$ $p_2$ ... -$through$ $p_n$ -$to$ $p_{n+1}$} \nonumber\\
&\text{\equationfontsize $set\_min\_delay$ $d^\prime$ -$from$ $p_1$ -$through$ $p_2$ ... -$through$ $p_n$ -$to$ $p_{n+1}$} \nonumber
\end{nospaceflalign}
where ``$set\_max\_delay$" and ``$set\_min\_delay$" are used to establish setup and hold time 
constraints 
 for long paths and short paths, respectively, 
``$through$" means propagating through, ``$from$'' and ``$to$'' mean leaving from and arriving at, $d$ and $d^\prime$ are delay values and $p_i$ is a pin name. 
``-$from$ $p_1$'' and ``-$to$ $p_{n+1}$'' can be removed in the constraints as long as the timing constraints are sufficient to identify required paths. 
In these timing constraints, setup time and hold time of flip-flops are subtracted and added automatically.  

In \figname~\ref{fig:remove_ff},
the path from F1 to F4 has three logic waves since it traverses through two removal locations
along it, namely the connection between $g1$/ZN and $g2$/A
as well as $g2$/ZN and $g_5$/A1.
The wave-pipelining constraints from F1 to F4 can be written as follows 
\begin{nospaceflalign}
&\text{\equationfontsize $set\_max\_delay$ 3$\cdot T$ -$through$ $g_1$/ZN -$through$ $g_2$/A -$through$ }\label{eq:setup} \nonumber\\
&\text{\equationfontsize $g2$/ZN -through $g_5$/A1}\\
&\text{\equationfontsize $set\_min\_delay$ 2$\cdot T$ -$through$ $g_1$/ZN -$through$ $g_2$/A -$through$ }\label{eq:hold} \nonumber\\ 
&\text{\equationfontsize $g2$/ZN -through $g_5$/A1.}
\end{nospaceflalign}

The timing constraints of the paths with two logic waves in \figname~\ref{fig:remove_ff}, such as 
the paths from F2 to F4 and from F3 to F4
can be 
written similarly. %
\begin{figure}[t]
{
\figurefontsize
\centering
\begingroup%
  \makeatletter%
  \providecommand\color[2][]{%
    \errmessage{(Inkscape) Color is used for the text in Inkscape, but the package 'color.sty' is not loaded}%
    \renewcommand\color[2][]{}%
  }%
  \providecommand\transparent[1]{%
    \errmessage{(Inkscape) Transparency is used (non-zero) for the text in Inkscape, but the package 'transparent.sty' is not loaded}%
    \renewcommand\transparent[1]{}%
  }%
  \providecommand\rotatebox[2]{#2}%
  \ifx\svgwidth\undefined%
    \setlength{\unitlength}{207.92816306bp}%
    \ifx\svgscale\undefined%
      \relax%
    \else%
      \setlength{\unitlength}{\unitlength * \real{\svgscale}}%
    \fi%
  \else%
    \setlength{\unitlength}{\svgwidth}%
  \fi%
  \global\let\svgwidth\undefined%
  \global\let\svgscale\undefined%
  \makeatother%
  \begin{picture}(1,0.66053384)%
    \put(0,0){\includegraphics[width=\unitlength,page=1]{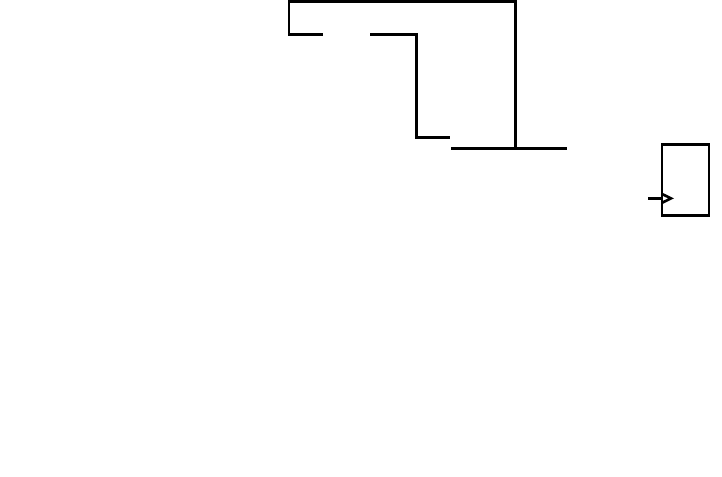}}%
    \put(0.94859415,0.39985259){\color[rgb]{0,0,0}\makebox(0,0)[b]{\smash{F5}}}%
    \put(0,0){\includegraphics[width=\unitlength,page=2]{separation1.pdf}}%
    \put(0.4795964,0.5744977){\color[rgb]{0,0,0}\makebox(0,0)[b]{\smash{F4}}}%
    \put(0,0){\includegraphics[width=\unitlength,page=3]{separation1.pdf}}%
    \put(0.05058108,0.55204687){\color[rgb]{0,0,0}\makebox(0,0)[b]{\smash{F1}}}%
    \put(0,0){\includegraphics[width=\unitlength,page=4]{separation1.pdf}}%
    \put(0.05058036,0.40163541){\color[rgb]{0,0,0}\makebox(0,0)[b]{\smash{F2}}}%
    \put(0,0){\includegraphics[width=\unitlength,page=5]{separation1.pdf}}%
    \put(0.3181416,0.46217316){\color[rgb]{1,0,0}\makebox(0,0)[lt]{\begin{minipage}{0.1466564\unitlength}\centering anchor \end{minipage}}}%
    \put(0,0){\includegraphics[width=\unitlength,page=6]{separation1.pdf}}%
    \put(0.05058036,0.19175212){\color[rgb]{0,0,0}\makebox(0,0)[b]{\smash{F3}}}%
    \put(0,0){\includegraphics[width=\unitlength,page=7]{separation1.pdf}}%
    \put(0.31784571,0.42087323){\color[rgb]{1,0,0}\makebox(0,0)[lt]{\begin{minipage}{0.1466564\unitlength}\centering points\end{minipage}}}%
    \put(0,0){\includegraphics[width=\unitlength,page=8]{separation1.pdf}}%
    \put(0.83853508,0.1271153){\color[rgb]{0,0,1}\makebox(0,0)[lt]{\begin{minipage}{0.32463134\unitlength}\raggedright 3 logic waves\end{minipage}}}%
    \put(0,0){\includegraphics[width=\unitlength,page=9]{separation1.pdf}}%
    \put(0.83856326,0.07661707){\color[rgb]{0,1,0}\makebox(0,0)[lt]{\begin{minipage}{0.32463134\unitlength}\raggedright 2 logic waves\end{minipage}}}%
    \put(0,0){\includegraphics[width=\unitlength,page=10]{separation1.pdf}}%
    \put(0.83856326,0.02611887){\color[rgb]{0,1,0}\makebox(0,0)[lt]{\begin{minipage}{0.32463134\unitlength}\raggedright 2 logic waves\end{minipage}}}%
    \put(0,0){\includegraphics[width=\unitlength,page=11]{separation1.pdf}}%
    \put(0.94816218,0.20221617){\color[rgb]{0,0,0}\makebox(0,0)[b]{\smash{F6}}}%
    \put(0,0){\includegraphics[width=\unitlength,page=12]{separation1.pdf}}%
  \end{picture}%
\endgroup%

\caption{Wave-pipelining paths with different number of logic waves are entangled with each other. Red dots represent anchor points, also the removal locations of flip-flops. $P1$, $P2$ and $P_3$ are the path set. The paths in $P_1$, as shown in blue, have three logic waves propagating along them and the paths in $P_2$ and $P_3$, as shown in green, have two logic waves propagating along them.}
\label{fig:separation}
}
\end{figure}

In \figname~\ref{fig:remove_ff}, 
the wave-pipelining paths with different number of logic waves are clearly separated. 
However, in some optimized circuits, wave-pipelining paths with different number of logic waves 
are entangled with each other. 
\figname~\ref{fig:separation} illustrates such a circuit, where 
the paths in $P_1$, as shown in blue color, %
has three logic waves propagating along it, 
since they traverse through two anchor points. 
The wave-pipelining constraints for such paths 
can be established with the two anchor points directly, similar to (\ref{eq:setup})-(\ref{eq:hold}).  
For example, the setup time constraint of such paths can be written as follows 
\begin{nospaceflalign}
&\text{\equationfontsize $set\_max\_delay$ 3$\cdot T$ -$through$ $g_1$/ZN -$through$ $g_4$/A1 -$through$ }\nonumber\\
&\text{\equationfontsize $g4$/ZN -$through$ $g_5$/A1}\label{eq:setup1} %
\end{nospaceflalign}
However, the paths in $P_2$ and $P_3$,  
as shown in green color, propagate through only one of the anchor points and 
entangle with the blue paths with three logic waves. 
Directly using one of the anchor points to  
establish the wave-pipelining constraints for such paths with two logic waves %
leads to a conflict with
the wave-pipelining constraints established with two anchor points. 
For example, if the setup constraint for the paths in $P_2$ is written as follow 
\begin{nospaceflalign}
&\text{\equationfontsize $set\_max\_delay$ 2$\cdot T$ -$through$ $g_1$/ZN -$through$ $g_4$/A1} \label{eq:setup2}
\end{nospaceflalign}
this constraint and the constraint (\ref{eq:setup1}) confuse Design Compiler, 
since the paths propagating through 
two anchor points also traverse through one of the anchor points. In this case, 
Design Compiler ignores the constraint in (\ref{eq:setup2}), so that the delays of the paths in $P_2$ 
cannot be optimized correctly. 
To solve this problem, 
the paths with two logic waves 
should be 
differentiated from the paths with three logic waves 
when establishing the corresponding timing constraints for the subsequent optimization. 

To differentiate the wave-pipelining paths with two logic waves from those with three logic waves,  
we propose 
to identify the pins that separate such paths. %
We call such pins \textbf{differentiating pins}.  
For example, 
to differentiate the wave-pipelining paths in $P_2$ from those in $P_1$, %
we first find the sink flip-flops at which the paths in $P_2$ and $P_1$ arrive. %
As shown in \figname~\ref{fig:separation},
F5 and F6 are such sink flip-flops. 
Since F6 is the sink flip-flop at which only 
the paths in $P_2$ can arrive, 
the wave-pipelining constraints in such a case can be established with the anchor point between  
$g_1$/ZN and $g_4$/A1 
together with the sink flip-flop as follows 
\begin{nospaceflalign} 
&\text{\equationfontsize $set\_max\_delay$ 2$\cdot T$ -$through$  $g_1$/ZN -$through$ $g_4$/A1 -$to$  F6/D}\nonumber\\
&\text{\equationfontsize $set\_min\_delay$ 1$\cdot T$ -$through$ $g_1$/ZN -$through$ $g_4$/A1 -$to$ F6/D.}\nonumber
\end{nospaceflalign}   

\begin{table*}[t]
\renewcommand{\tabcolsep}{4.15pt}
\centering
{\small
\renewcommand{\arraystretch}{1}

\vskip 10pt
\caption{Results of \titlenym}
\vskip -2pt
\label{tb_test}
\begin{tabular}{lrrr r rrrr r rr r rrr r rr r r} \hlinewd{0.7pt}

\multicolumn{4}{c}{Circuit} &

\multicolumn{1}{c}{} & 

\multicolumn{4}{c}{Ret.\&Siz.} &

\multicolumn{1}{c}{} &

\multicolumn{2}{c}{Cri. Part} &

\multicolumn{1}{c}{} &

\multicolumn{3}{c}{Opt. Circuit} &

\multicolumn{1}{c}{} &

\multicolumn{2}{c}{Comparison} &

\multicolumn{1}{c}{} &

\multicolumn{1}{c}{Runtime}  \\

\cline {1-4} \cline{6-9}  \cline{11-12}  \cline{14-16} \cline {18-19} \cline {21-21}

\multicolumn{1}{c}{} &
\multicolumn{1}{c}{$n_s$} &
\multicolumn{1}{c}{$n_c$} &
\multicolumn{1}{c}{$T(ps)$} &

\multicolumn{1}{c}{} &

\multicolumn{1}{c}{$n^\prime_{s}$} &
\multicolumn{1}{c}{$n^\prime_{c}$} &
\multicolumn{1}{c}{$n^\prime_{t}$} &
\multicolumn{1}{c}{$n^\prime_{a}$} &
\multicolumn{1}{c}{} &

\multicolumn{1}{c}{$n_{cs}$} & 
\multicolumn{1}{c}{$n_{cc}$} & 

\multicolumn{1}{c}{} &

\multicolumn{1}{c}{$n_{f}$} & 
\multicolumn{1}{c}{$n_{l}$} &
\multicolumn{1}{c}{$n_{b}$} &

\multicolumn{1}{c}{} &

\multicolumn{1}{c}{$n_{t}$} &
\multicolumn{1}{c}{$n_{a}$} &

\multicolumn{1}{c}{} &

\multicolumn{1}{c}{$t(s)$} \\

\hlinewd{0.8pt}

systemcdes &574  &1286 &1132 & &523 &2429 &55.5\% &23.1\% & &195  &2290 &  &65 &66  & 31& &2.5\%  &-5.4\%  & &2619.5 \\
tv80     &1014 &2998 &1025 & &1228 &5818 &40.0\% &50.4\% & &296 &5758 &  &158 &141  & 383 & &2\%  &0.6\% & &3081.5 \\
wb\_dma     &720  &1632 &666 & &716 &1974 &31.5\% &11.4\% & &129  &1375 &  &65 &46  & 240 & &8.5\%    &1.5\%  & &2211.3 \\
systemcaes &1181  &3318 &921 & &1635 &5078 &29.5\% &38.0\% & &363  &4531 &  &258 &150  & 141 & &3\%    &-1.5\%  & &2052.1 \\
mem\_ctrl  &879 &1840 &1105 & &1429 &3318 &51.0\% &68.3\% & &290 &2642 &  &156&93  & 296& &4.5\%  &-1.5\%  & &4510.5 \\
usb\_funct &2699 &4952 &1042 & &3005 &8901 &35.5\% &28.6\% & &333 &6293 &  &253&71  & 252 & &3\%    &0.1\%  & &2166.5 \\
ac97\_ctrl &1145 &1467 &802 & &2021 &2363 &48.5\% &70.9\% & &579 &1978 &  &430 &125 & 163& &1\%    &-2.2\% & &1961.2 \\
pci\_bridge&2301 &4993 &1141 & &6243 &6476 &33.5\% &130.1\% & &264 &4410 &  &156&48  & 140& &1\%    &-0.7\%  & &3563.1 \\

\hlinewd{0.7pt}
\end{tabular}
}
\end{table*}

Since F5 is the sink flip-flop at which both wave-pipelining paths in $P_1$ and $P_2$ can arrive, 
we need to find the differentiating pins to separate them. 
To achieve this goal, 
we enumerate 
the paths starting from F5 and ending at the two anchor points backwards.   
For example, the path propagating through $g6$/A2, $g_3$/A, and $g_4$/A1 terminates at 
the anchor point between $g_1$/ZN and $g_4$/A1 
and 
the path propagating through $g6$/A1 and $g5$/A1 terminates at the anchor point between $g_4$/ZN and $g_5$/A1. 
By comparing these paths, it is clear that 
the pin $g_6$/A2 differentiates the wave-pipelining paths in $P_2$ from those in $P_1$. 
Therefore, the wave-pipelining constraints for the paths in $P_2$ can be established with the 
differentiating pin $g_6$/A2 and the anchor point between $g_1$/ZN and $g_4$/A1 
as follows
\begingroup
\allowdisplaybreaks
\begin{nospaceflalign}  
&\text{\equationfontsize $set\_max\_delay$ 2$\cdot T$ -$through$ $g_1$/ZN -$through$ $g_4$/A1 -$through$ } \nonumber\\
&\text{\equationfontsize  $g_6$/A2}\nonumber\\
&\text{\equationfontsize $set\_min\_delay$ 1$\cdot T$ -$through$ $g_1$/ZN -$through$ $g_4$/A1 -$through$ } \nonumber\\
&\text{\equationfontsize  $g_6$/A2.}\nonumber
\end{nospaceflalign} 
\endgroup
Similarly, to differentiate the wave-pipelining paths in $P_3$ from those in $P_1$, we 
first find the common source flip-flops which the paths in $P_3$ and $P_1$ start from. 
As shown in \figname~\ref{fig:separation}, this common source flip-flop is F2. 
Afterwards, we enumerate the paths starting 
from F2 and ending at the two anchor points in a forward way. 
By comparing these paths, 
we can see that the pin $g_2$/A1 differentiates 
the wave-pipelining paths in $P_3$ from those in $P_1$. 
We then use the differentiating pin $g_2$/A1 and the anchor point between $g_4$/ZN and $g_5$/A1 to establish the wave-pipelining constraints. 
In a general case where $N$ anchor points exist in an optimized circuit after 
the removal locations of flip-flops are extracted, %
we use the methods described above 
to establish the wave-pipelining constraints for the 
paths where the number of logic waves propagating along them is larger than 1. 
The generation of wave-pipelining constraints is fully automated without manual intervention. 

\subsection{Circuit Optimizing and Buffer Insertion with Commercial Tools}\label{sec:fourth_inte}

After the wave-pipelining constraints are established for the optimized circuit, 
we 
incorporate such constraints into Design Compiler to optimize this circuit. 
During the optimization,  
Design Compiler sizes gates to increase the delays in short paths to satisfy 
the wave-pipelining constraints. 
After this optimization, the delays of the short paths might still be 
too small to satisfy the wave-pipelining constraints. To pad their delays with Design Compiler, %
we insert buffers in the optimized circuits to enlarge their delays with 
the command ``insert\_buffer $p$ -no\_of\_cells $d$ $t$", where $p$ is a pin name, $d$ is the number of buffers and $t$ is the type of 
buffers, e.g., BUF\_X1.  
The locations of buffers can be obtained from the adjusted iterative relaxation. Due 
to the heuristic optimization in the adjusted iterative relaxation, 
the number of buffers might not be large or small enough to satisfy the wave-pipelining constraints. 
To solve this problem, 
we iteratively insert/remove buffers along the short/long paths until the 
wave-pipelining constraints are met.

\section{Experimental Results}\label{sec:results}

\subsection{Experimental Setup}

The proposed method was implemented in C++ and tested
using a \SI[mode=text]{3.20}{\GHz} CPU. 
In the experiments, 
the circuits from the TAU 2013 variation-aware timing analysis 
contest are first optimized using \SI[mode=text]{45}{\nm} library with Design Compiler to reduce area cost. 
These circuits
are referred to the original circuits. 
\reviewhl{The clock periods of these original circuits are shown as $T$ in the fourth column.}
We optimize these circuits further using the proposed \titlenym\ to 
improve timing performance and reduce area. The optimization results are shown in Table~\ref{tb_test}. 
The number of flip-flops and the number of combinational components of the original circuits are shown in the  
columns $\boldsymbol{n_s}$ and $\boldsymbol{n_c}$, respectively.
To tolerate process variations, 10\% of timing margin was assigned, so that $r^u$
and $r^l$ in Section~\ref{sec:impl} were set to 1.1 and 0.9, respectively. 
\reviewhl{In case of large variations in advanced technologies, a large timing margin can be 
assigned by increasing $r^u$ and decreasing $r^l$, respectively.}
The allowed phase shifts $\phi_{wt}$ in Section~\ref{sec:impl} are 0, T/4, T/2 and 3T/4. 
\reviewhl{$t_{stable}$ in equation (17) in Section~\ref{sec:model} is set to the delay of a buffer 
to isolate signals on the next wave and the previous wave. 
The initial delay bound $d_{th}$ in Section~\ref{sec:impl} is 7T/8 and iteratively reduced by T/8.}
The ILP solver used in the \titlenym\ framework was Gurobi.

\reviewhl{%
The goal of VirtualSync+ is to enhance timing performance and thus 
solve timing violations in the circuits
which have already been optimized.  
Therefore, 
the original circuits 
are first retimed and sized extremely %
with Design Compiler to achieve the limit of timing performance. 
Accordingly, extreme retiming and sizing is an integral part of the VirtualSync+ framework. 
}
This extreme retiming and sizing, abbreviated as extreme R\&S, 
is realized by iteratively reducing the clock period and optimizing the circuits with retiming and sizing in Design Compiler
until the circuits cannot be optimized further anymore. 
In the circuits after extreme R\&S, 
the number of flip-flops and the number of combinational components, shown as $\boldsymbol{n^\prime_{s}}$ and $\boldsymbol{n^\prime_{c}}$ in
Table~\ref{tb_test}, are different from those in the original circuits. 
$\boldsymbol{n^\prime_t}$ and $\boldsymbol{n^\prime_c}$ show the clock period reduction 
and the area increase of the circuits after extreme R\&S with respect to the original circuits. 
Overall, the timing performance 
of a circuit is improved significantly by extreme R\&S while the area is increased. 
\begin{figure*}
\figurefontsize

\begin{minipage}[t]{0.32\textwidth}%
\centering
\pgfplotsset{compat=1.3,
    /pgfplots/ybar legend/.append style={ 
        /pgfplots/legend image code/.code={%
           \draw[##1,/tikz/.cd,yshift=-0.25em]
           (0cm,0cm) rectangle (7pt,0.8em);
        },
    }
}

\begin{tikzpicture}

\pgfplotstableread[row sep=\\]{
circuit  Original  New\\
1        131        50\\
2        299        193\\
3        111        62\\
4        408        302\\
5        249       120\\
6        324       196\\
7        555       313\\
8       204       125\\
}\loadedtable

\begin{axis}[
xticklabels={systemcdes, tv80, wb\_dma, systemcaes,mem\_ctrl, usb\_funct, ac97\_ctr, pci\_bridge}, 
xtick={1,...,8},
xmin=0.3, xmax=8.8,
x=0.63cm, y=0.003188406cm, 
x tick label style={rotate=45, xshift=0pt,yshift=0pt,anchor=east, 
inner sep=0}, 
xticklabel pos=left, xtick align=outside, xtick pos=left,
ymin=0, ymax=690,
ylabel={Num.of seq. units}, 
ylabel style={inner sep=0}, 
ylabel shift=0pt, ytickmin=0,ytickmax=690, 
legend columns=3, 
legend style={
at={(0.40,0.850)}, anchor=center, 
/tikz/every even column/.append style={column sep=0.2cm},
draw=none, 
},
line width=0.75pt,
ybar=0pt, 
bar width=6pt,
axis on top=true,
major tick length=3pt,
]  
\addplot[ybar, line width=0.35pt, red, fill=red!30!white] table[x=circuit,y=Original] {\loadedtable};
\addplot[ybar, line width=0.35pt, blue, fill=blue!30!white] table[x=circuit,y=New] {\loadedtable};
\legend{w/ rep., w/o rep.}

\end{axis}
\end{tikzpicture}

\vskip  -2.0pt
\caption{Comparison of sequential delay units after buffer replacement.}\label{fig:se_cmp}
\end{minipage}
~\hspace{5pt}
\begin{minipage}[t]{0.32\textwidth}%
\centering
\pgfplotsset{compat=1.3,
    /pgfplots/ybar legend/.append style={ 
        /pgfplots/legend image code/.code={%
           \draw[##1,/tikz/.cd,yshift=-0.25em]
           (0cm,0cm) rectangle (5pt,0.8em);
        },
    }
}

\begin{tikzpicture}

\pgfplotstableread[row sep=\\]{
circuit ideal\\
1       0.673\\
2       0.607\\
3       0.731\\
4       0.521\\
5       0.427\\
6       0.576\\
7       0.459\\
8      0.774\\
}\loadedtable

\begin{axis}[
xticklabels={systemcdes, tv80, wb\_dma, systemcaes, mem\_ctrl, usb\_funct, ac97\_ctr, pci\_bridge},
xtick={1,...,8},
xmin=1, xmax=8.0,
x=0.625cm, y=2.2cm, 
x tick label style={rotate=45, xshift=0pt,yshift=0pt,anchor=east, 
inner sep=0}, 
xticklabel pos=left, xtick align=outside, xtick pos=left,
ymin=0, ymax=1, 
ylabel={Area ratio }, 
ylabel style={inner sep=0}, 
ylabel shift=0pt, ytickmin=0,ytickmax=20, 
legend columns=3, 
legend style={
at={(0.5,0.87)}, anchor=center, 
/tikz/every even column/.append style={column sep=0.2cm},
draw=none, 
},
line width=0.75pt,
major tick length=3pt,
]  \addplot[sharp plot, line width=0.5pt, cyan!60!blue, mark=o, fill=none] table[x=circuit,y=ideal] {\loadedtable};
\end{axis}
\end{tikzpicture}

\vskip -2pt
\caption{Area comparison before and after buffer replacement.}\label{fig:buffer_replace}
\end{minipage}
~\hspace{5pt}
\begin{minipage}[t]{0.32\textwidth}%
\centering
\pgfplotsset{scaled y ticks=false}
\pgfplotsset{compat=1.3,
    /pgfplots/ybar legend/.append style={ 
        /pgfplots/legend image code/.code={%
           \draw[##1,/tikz/.cd,yshift=-0.25em]
           (0cm,0cm) rectangle (7pt,0.8em);
        },
    }
}

\begin{tikzpicture}

\pgfplotstableread[row sep=\\]{
circuit  Original  New\\
1        0.69445     0.65156\\
2        1.16928     1.16581\\
3        0.56097      0.56042\\
4        1.41778      1.38927\\
5        1.04748      1.02792\\
6        2.25698      2.2573\\
7        1.17974      1.11256\\
8        3.76836     3.65147\\
}\loadedtable

\begin{axis}[
scaled y ticks = false,
xticklabels={systemcdes, tv80, wb\_dma, systemcaes, mem\_ctrl, usb\_funct, ac97\_ctr, pci\_bridge}, 
xtick={1,...,8},
xmin=0.3, xmax=8.8,
x=0.63cm, y=5.5mm,
x tick label style={rotate=45, xshift=0pt,yshift=0pt,anchor=east, 
inner sep=0}, 
xticklabel pos=left, xtick align=outside, xtick pos=left,
ymin=0, ymax=4, 
ylabel={Area cmp. ($\times 10^4$$\upmu m^{2}$)}, 
ylabel style={inner sep=0}, 
ylabel shift=-2pt, ytickmin=0,ytickmax=4, 
legend columns=3, 
legend style={
at={(0.45,0.842)}, anchor=center, 
/tikz/every even column/.append style={column sep=0.2cm},
draw=none, 
},
line width=0.75pt,
ybar=0pt, 
bar width=6pt,
axis on top=true,
major tick length=3pt,
]  
\addplot[ybar, line width=0.35pt, red, fill=red!30!white] table[x=circuit,y=Original] {\loadedtable};
\addplot[ybar, line width=0.35pt, blue, fill=blue!30!white] table[x=circuit,y=New] {\loadedtable};
\legend{Retiming, \titlenym}

\end{axis}
\end{tikzpicture}

\vskip -2pt
\caption{Area comparisons with extreme R\&S with the same clock period.}\label{fig:samet_area_cmp}
\end{minipage}
\end{figure*}

To increase the timing performance with \titlenym, 
in the circuits after extreme R\&S, 
combinational paths whose delays are larger than a specified clock 
period, which is evaluated by iteratively reducing the clock period obtained from extreme R\&S,  
were selected. 
The source and sink flip-flops of these selected paths were allowed to be
removed, while the other flip-flops in the circuits were
considered as boundary flip-flops. 
All the combinational components that can
reach the flip-flops at the sources or sinks of these selected paths 
through combinational paths 
are considered as the critical part of a circuit together. 
The extracted critical part of the circuits 
occupied a large portion of the circuits after extreme R\&S, 
as shown in %
$\boldsymbol{n_{cs}}$ and $\boldsymbol{n_{cc}}$, which 
represent the percentage of the number of flip-flops and combinational components 
of the critical part in the circuits after extreme R\&S. 
From $\boldsymbol{n_{cs}}$ and $\boldsymbol{n_{cc}}$, we can see that 
more than 4\% of flip-flops and more than 68\% of combinational components have
been selected for timing optimization.
\reviewhl{
In case of industrial designs which contain
hundreds of thousands of
gates in the extracted critical subcircuit, 
the critical subcircuit can be processed with VirtualSync+ by iteratively selecting a small part for timing optimization.
}

\begin{figure}[h]
{
\figurefontsize
\centering
\pgfplotsset{compat=1.3,
    /pgfplots/ybar legend/.append style={ 
        /pgfplots/legend image code/.code={%
           \draw[##1,/tikz/.cd,yshift=-0.25em]
           (0cm,0cm) rectangle (5pt,0.8em);
        },
    }
}

\begin{tikzpicture}

\pgfplotstableread[row sep=\\]{
circuit ideal real\\  
1       2.5   1\\ 
2       2     0\\
3       8.5   4\\
4       3     3\\
5       4.5  2.5\\
6       3    1\\
7       1    0.5\\
8       1    0\\
}\loadedtable

\begin{axis}[
xticklabels={systemcdes, tv80, wb\_dma, systemcaes, mem\_ctrl, usb\_funct, ac97\_ctr, pci\_bridge},
xtick={1,...,8},
xmin=1, xmax=8.0,
x=0.825cm, y=0.22cm, 
x tick label style={rotate=45, xshift=0pt,yshift=0pt,anchor=east, 
inner sep=0}, 
xticklabel pos=left, xtick align=outside, xtick pos=left,
ymin=0, ymax=11.5, 
ylabel style={align=center}, ylabel=clock period reduction\\compared with\\extreme R\&S (\%),
ylabel style={inner sep=0}, 
ylabel shift=0pt, ytickmin=0,ytickmax=20, 
legend columns=3, 
legend style={
at={(0.5,0.87)}, anchor=center, 
/tikz/every even column/.append style={column sep=0.2cm},
draw=none, 
},
line width=0.75pt,
major tick length=3pt,
]  \addplot[sharp plot, line width=0.5pt, cyan!60!blue, mark=o, fill=none] table[x=circuit,y=ideal] {\loadedtable};
   \addplot[sharp plot, line width=0.5pt, green!50!black, mark=triangle, fill=none] table[x=circuit,y=real] {\loadedtable};
   \legend{w/o fine-tuning, w/ fine-tuning}
\end{axis}
\end{tikzpicture}

\caption{Comparison of the clock period reduction before and after fine-tuning with Design Compiler.}
\label{fig:timing_com}
}
\end{figure}
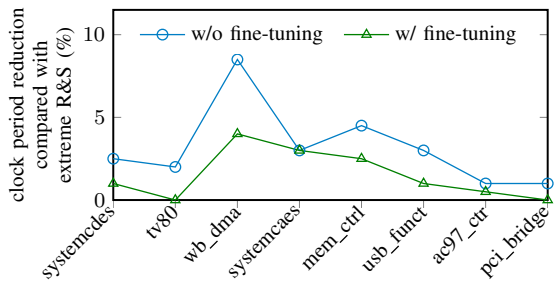

\begin{figure}[h]
{
\figurefontsize
\centering
\pgfplotsset{compat=1.3,
    /pgfplots/ybar legend/.append style={ 
        /pgfplots/legend image code/.code={%
           \draw[##1,/tikz/.cd,yshift=-0.25em]
           (0cm,0cm) rectangle (5pt,0.8em);
        },
    }
}

\begin{tikzpicture}

\pgfplotstableread[row sep=\\]{
circuit ideal real\\  
1       -5.4   1\\ 
2       0.6    -1.9\\
3       1.5    4\\
4       -1.5    -0.4\\
5       -1.5  -7\\
6       0.1    -1\\
7       -2.2    -11.5\\
8       -0.7    -2.8\\
}\loadedtable

\begin{axis}[
xticklabels={systemcdes, tv80, wb\_dma, systemcaes, mem\_ctrl, usb\_funct, ac97\_ctr, pci\_bridge},
xtick={1,...,8},
xmin=1, xmax=8.0,
x=0.825cm, y=0.1205cm, 
x tick label style={rotate=45, xshift=0pt,yshift=0pt,anchor=east, 
inner sep=0}, 
xticklabel pos=left, xtick align=outside, xtick pos=left,
ymin=-12, ymax=9, 
ylabel style={align=center}, ylabel=area increase\\compared with\\extreme R\&S (\%),
ylabel style={inner sep=0}, 
ylabel shift=0pt, ytickmin=-12,ytickmax=9, 
legend columns=3, 
legend style={
at={(0.5,0.89)}, anchor=center, 
/tikz/every even column/.append style={column sep=0.2cm},
draw=none, 
},
line width=0.75pt,
major tick length=3pt,
]  \addplot[sharp plot, line width=0.5pt, cyan!60!blue, mark=o, fill=none] table[x=circuit,y=ideal] {\loadedtable};
   \addplot[sharp plot, line width=0.5pt, green!50!black, mark=triangle, fill=none] table[x=circuit,y=real] {\loadedtable};
   \legend{w/o fine-tuning, w/ fine-tuning}
\end{axis}
\end{tikzpicture}

\caption{Comparison of the area increase before and after fine-tuning with Design Compiler.}
\label{fig:area_com}
}
\end{figure}
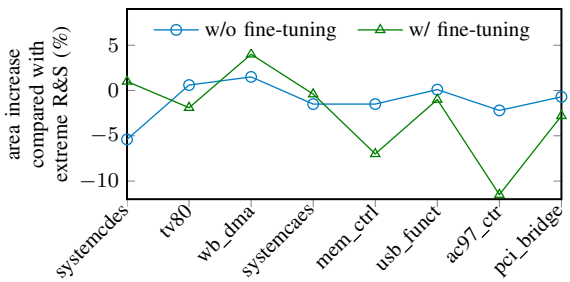

\subsection{Experimental Results with Timing Optimization Using Relaxed \titlenym\ Timing Model}\label{sec:secondstep}

To verify the improvement of circuit performance, we gradually reduced the
clock period by 0.5\% of the clock period obtained from extreme R\&S
and applied the timing optimization with relaxed \titlenym\ in Section~\ref{sec:impl} 
to meet the timing constraints. %
The results correspond to the results described 
in the conference version \cite{Grace2018_DAC} of this paper. %
The column $\boldsymbol{n_f}$ and $\boldsymbol{n_l}$ %
show the numbers of flip-flops and latches in the critical part after optimization with relaxed \titlenym, respectively.
The sums of these numbers are comparable or even smaller than the numbers of
flip-flops before relaxed \titlenym.  %
The numbers of extra inserted buffers to match arrival times 
are shown in the column $\boldsymbol{n_{b}}$. 
Thanks to the buffer replacement with sequential units in the proposed framework,
the numbers of extra inserted are not large. %
Compared with the number of original combinational components shown in the column $n_{cc}$, 
these numbers show that the cost due to the inserted buffers is still acceptable.

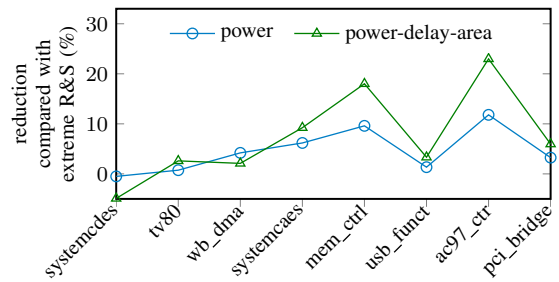
\begin{figure}[t]
{
\figurefontsize
\centering
\pgfplotsset{compat=1.3,
    /pgfplots/ybar legend/.append style={ 
        /pgfplots/legend image code/.code={%
           \draw[##1,/tikz/.cd,yshift=-0.25em]
           (0cm,0cm) rectangle (5pt,0.8em);
        },
    }
}

\begin{tikzpicture}

\pgfplotstableread[row sep=\\]{
circuit ideal real\\  
1       -0.5   -4.9\\ 
2       0.74     2.6\\
3       4.2     2.1\\
4       6.18     9.23\\
5       9.6     18.01\\
6       1.34    3.33\\
7       11.8    22.98\\
8       3.25    5.96\\
}\loadedtable

\begin{axis}[
xticklabels={systemcdes, tv80, wb\_dma, systemcaes, mem\_ctrl, usb\_funct, ac97\_ctr, pci\_bridge},
xtick={1,...,8},
xmin=1, xmax=8.0,
x=0.825cm, y=0.06657895cm, 
x tick label style={rotate=45, xshift=0pt,yshift=0pt,anchor=east, 
inner sep=0}, 
xticklabel pos=left, xtick align=outside, xtick pos=left,
ymin=-5, ymax=33, 
ylabel style={align=center}, ylabel=reduction\\compared with\\extreme R\&S (\%),
ylabel style={inner sep=0}, 
ylabel shift=0pt, ytickmin=-5,ytickmax=33, 
legend columns=3, 
legend style={
at={(0.5,0.87)}, anchor=center, 
/tikz/every even column/.append style={column sep=0.2cm},
draw=none, 
},
line width=0.75pt,
major tick length=3pt,
]  \addplot[sharp plot, line width=0.5pt, cyan!60!blue, mark=o, fill=none] table[x=circuit,y=ideal] {\loadedtable};
   \addplot[sharp plot, line width=0.5pt, green!50!black, mark=triangle, fill=none] table[x=circuit,y=real] {\loadedtable};
   \legend{power, power-delay-area}
\end{axis}
\end{tikzpicture}
  
\caption{The reduction of power and power-delay-area product compared with extreme R\&S.}
\label{fig:power_com}
}
\end{figure}

\begin{figure*}
\begin{minipage}{1\textwidth}%
\figurefontsize
\centering
\subfloat[][pci\_bridge]{\makebox{\pgfplotsset{compat=1.3,
    /pgfplots/ybar legend/.append style={ 
        /pgfplots/legend image code/.code={%
           \draw[##1,/tikz/.cd,yshift=-0.25em]
           (0cm,0cm) rectangle (5pt,0.8em);
        },
    }
}

\begin{tikzpicture}

\pgfplotstableread[row sep=\\]{
gap   re\\
1     37683.4\\
2     nan \\
3     26911\\
4     nan\\
5     21948.5\\
6     nan\\ 
7     21397.6\\
8     nan\\
9     21077.6\\
10    nan\\
11    20948.0\\
}\loadedtable

\pgfplotstableread[row sep=\\]{
gap    our\\     
1     36629.5\\
2     25233\\
3     nan\\
4     21142.6\\
5     nan\\
6     20999.5\\
7     nan\\
8     20973.1\\
9     nan\\
10    20950.1\\
11    nan\\
}\loadedtable

\begin{axis}[ 
scatter/classes={
a={mark=x,blue,solid},%
b={mark=triangle*,red,solid},%
c={mark=o,draw=black,solid},
d={mark=halfsquare*,green,solid},
e={mark=*,yellow,solid},
f={mark=star,purple,solid}},
xticklabels={1T, 1.05T, 1.10T,1.15T, 1.20T, 1.25T},
xtick={1, 1.05, 1.10,1.15, 1.20, 1.25},
xmin=1, xmax=1.25,
x=13.562cm, y=0.0001265cm, 
x tick label style={rotate=0, xshift=0pt,yshift=-3pt,anchor=east, font=\fontsize{6.5}{6.5}\selectfont,
inner sep=0}, 
xticklabel pos=left, xtick align=outside, xtick pos=left,
ymin=20000, ymax=40000, 
ylabel={Area ($\upmu m^{2}$)}, 
y tick label style={ font=\fontsize{7}{7}\selectfont},
ylabel style={inner sep=0, font=\fontsize{7}{7}\selectfont}, 
ylabel shift=-2pt, ytickmin=25000,ytickmax=40000, 
legend columns=1, 
legend style={
at={(0.53,0.81)}, anchor=center, 
/tikz/every even column/.append style={column sep=0.05cm},
draw=none, 
},
line width=0.55pt,
major tick length=2.5pt,
legend style={font=\fontsize{7}{7}\selectfont},
legend entries={{Retiming\&sizing},{VirtualSync+}},
] 
\addlegendimage{no markers, black,dashed}
\addlegendimage{no markers, blue}
\addplot[scatter, line width=0.5pt, black, dashed, dash pattern=on 3pt off 3pt,dash phase=8pt, domain = -5:5, 
scatter src=explicit symbolic]
coordinates {
(1,37683.4) [a]
(1.05,26911) [b]
(1.1,21948.5) [c]
(1.15,21397.6) [d]
(1.20,21077.6) [e]
(1.25,20948.0) [f]
};

\addplot[scatter, line width=0.5pt, blue, 
scatter src=explicit symbolic]
coordinates {
(1,36629.5) [a]
(1.03,25233) [b]
(1.08,21142.6) [c]
(1.12,20999.5) [d]
(1.17,20973.1) [e]
(1.23,20950.1) [f]
};

\end{axis}
\end{tikzpicture}
}
}%
\subfloat[][ac97\_ctrl]{\makebox{\pgfplotsset{compat=1.3,
    /pgfplots/ybar legend/.append style={ 
        /pgfplots/legend image code/.code={%
           \draw[##1,/tikz/.cd,yshift=-0.25em]
           (0cm,0cm) rectangle (5pt,0.8em);
        },
    }
}

\begin{tikzpicture}
\pgfplotstableread[row sep=\\]{
gap   re\\
1     37683.4\\
2     nan\\
3     26911\\
4     nan\\
5     21948.5\\
6     nan\\ 
7     21397.6\\
8     nan\\
9     21077.6\\
10    nan\\
11    20948.0\\
}\loadedtable

\pgfplotstableread[row sep=\\]{
gap    our\\     
1     36629.5\\
2     25233\\
3     nan\\
4     21142.6\\
5     nan\\
6     20999.5\\
7     nan\\
8     20973.1\\
9     nan\\
10    20950.1\\
11    nan\\
}\loadedtable

\begin{axis}[ 
scatter/classes={
a={mark=x,blue,solid},%
b={mark=triangle*,red,solid},%
c={mark=o,draw=black,solid},
d={mark=halfsquare*,green,solid},
e={mark=*,yellow,solid},
f={mark=star,purple,solid}},
xticklabels={0.99T, 1.042T, 1.094T,1.146T, 1.198T, 1.25T},
xtick={0.99, 1.042, 1.094,1.146, 1.198, 1.25},
xmin=0.99, xmax=1.25,
x=13.562cm, y=0.000977cm, 
x tick label style={rotate=0, xshift=0pt,yshift=-3pt,anchor=east, font=\fontsize{6.5}{6.5}\selectfont,  
inner sep=0}, 
xticklabel pos=left, xtick align=outside, xtick pos=left,
ymin=9500, ymax=12000, 
ylabel={Area ($\upmu m^{2}$) }, 
y tick label style={ font=\fontsize{7}{7}\selectfont},
ylabel style={inner sep=0, font=\fontsize{7}{7}\selectfont}, 
ylabel shift=-2pt, ytickmin=9500,ytickmax=12000, 
legend columns=1, 
legend style={
font=\fontsize{7}{7}\selectfont,line width=0.5pt,
at={(0.55,0.81)}, anchor=center, 
/tikz/every even column/.append style={column sep=0.05cm},
draw=none, 
},
line width=0.55pt,
major tick length=2.5pt,
legend entries={{Retiming\&sizing},{VirtualSync+}},
] 
\addlegendimage{no markers, black,dashed}
\addlegendimage{no markers, blue}
\addplot[scatter, line width=0.5pt, black, dashed, dash pattern=on 3pt off 3pt,dash phase=8pt, domain = -5:5, 
scatter src=explicit symbolic]
coordinates {
(1,11797.4) [a]
(1.05,10017) [b]
(1.1,9862.7) [c]
(1.15,9860.0) [d]
(1.20,9808.0) [e]
(1.25,9765.9) [f]
};

\addplot[scatter, line width=0.5pt, blue, 
scatter src=explicit symbolic]
coordinates {
(0.99,10442.6) [a]
(1.04,9967.6) [b]
(1.09,9741.8) [c]
(1.14,9713.2) [d]
(1.19,9731.8) [e]
(1.23,9701.8) [f]
};

\end{axis}
\end{tikzpicture}
}
}
\subfloat[][usb\_funct]{\makebox{\pgfplotsset{compat=1.3,
    /pgfplots/ybar legend/.append style={ 
        /pgfplots/legend image code/.code={%
           \draw[##1,/tikz/.cd,yshift=-0.25em]
           (0cm,0cm) rectangle (5pt,0.8em);
        },
    }
}

\begin{tikzpicture}
\pgfplotstableread[row sep=\\]{
gap   re\\
1     37683.4\\
2     nan\\
3     26911\\
4     nan\\
5     21948.5\\
6     nan\\ 
7     21397.6\\
8     nan\\
9     21077.6\\
10    nan\\
11    20948.0\\
}\loadedtable

\pgfplotstableread[row sep=\\]{
gap    our\\     
1     36629.5\\
2     25233\\
3     nan\\
4     21142.6\\
5     nan\\
6     20999.5\\
7     nan\\
8     20973.1\\
9     nan\\
10    20950.1\\
11    nan\\
}\loadedtable

\begin{axis}[ 
scatter/classes={
a={mark=x,blue,solid},%
b={mark=triangle*,red,solid},%
c={mark=o,draw=black,solid},
d={mark=halfsquare*,green,solid},
e={mark=*,yellow,solid},
f={mark=star,purple,solid}},
xticklabels={0.99T, 1.042T, 1.094T,1.146T, 1.198T, 1.25T},
xtick={0.99, 1.042, 1.094,1.146, 1.198, 1.25},
xmin=0.99, xmax=1.25,
x=13.562cm, y=0.0006763cm, 
x tick label style={rotate=0, xshift=0pt,yshift=-3pt,anchor=east, font=\fontsize{6.5}{6.5}\selectfont,  
inner sep=0}, 
xticklabel pos=left, xtick align=outside, xtick pos=left,
ymin=20000, ymax=23600, 
ylabel={Area ($\upmu m^{2}$)}, 
y tick label style={ font=\fontsize{7}{7}\selectfont},
ylabel style={inner sep=0, font=\fontsize{7}{7}\selectfont}, 
ylabel shift=-2pt, ytickmin=21000,ytickmax=23600, 
legend columns=1, 
legend style={
font=\fontsize{7}{7}\selectfont,line width=0.5pt,
at={(0.55,0.81)}, anchor=center, 
/tikz/every even column/.append style={column sep=0.05cm},
draw=none, 
},
line width=0.55pt,
major tick length=2.5pt,
legend entries={{Retiming\&sizing},{VirtualSync+}},
] 
\addlegendimage{no markers, black,dashed}
\addlegendimage{no markers, blue}
\addplot[scatter, line width=0.5pt, black, dashed, dash pattern=on 3pt off 3pt,dash phase=8pt, domain = -5:5, 
scatter src=explicit symbolic]
coordinates {
(1,22569.8) [a]
(1.05,22213.4) [b]
(1.1,21875.8) [c]
(1.15,20660.8) [d]
(1.20,20606.2) [e]
(1.25,20444.2) [f]
};

\addplot[scatter, line width=0.5pt, blue, 
scatter src=explicit symbolic]
coordinates {
(0.99,22337.3) [a]
(1.03,22100.5) [b]
(1.08,21699.7) [c]
(1.13,20451.9) [d]
(1.18,20388.0) [e]
(1.22,20200.1) [f]
};

\end{axis}
\end{tikzpicture}
}
}
\subfloat[][mem\_ctrl]{\makebox{\pgfplotsset{compat=1.3,
    /pgfplots/ybar legend/.append style={ 
        /pgfplots/legend image code/.code={%
           \draw[##1,/tikz/.cd,yshift=-0.25em]
           (0cm,0cm) rectangle (5pt,0.8em);
        },
    }
}

\begin{tikzpicture}
\pgfplotstableread[row sep=\\]{
gap   re\\
1     37683.4\\
2     nan\\
3     26911\\
4     nan\\
5     21948.5\\
6     nan\\ 
7     21397.6\\
8     nan\\
9     21077.6\\
10    nan\\
11    20948.0\\
}\loadedtable

\pgfplotstableread[row sep=\\]{
gap    our\\     
1     36629.5\\
2     25233\\
3     nan\\
4     21142.6\\
5     nan\\
6     20999.5\\
7     nan\\
8     20973.1\\
9     nan\\
10    20950.1\\
11    nan\\
}\loadedtable

\begin{axis}[ 
scatter/classes={
a={mark=x,blue,solid},%
b={mark=triangle*,red,solid},%
c={mark=o,draw=black,solid},
d={mark=halfsquare*,green,solid},
e={mark=*,yellow,solid},
f={mark=star,purple,solid}},
xticklabels={0.97T, 1.026T, 1.082T,1.138T, 1.194T, 1.25T},
xtick={0.97, 1.026, 1.082,1.138, 1.194, 1.25},
xmin=0.97, xmax=1.25,
x=12.594cm, y=0.0006943cm, 
x tick label style={rotate=0, xshift=0pt,yshift=-3pt,anchor=east, font=\fontsize{6.5}{6.5}\selectfont,  
inner sep=0}, 
xticklabel pos=left, xtick align=outside, xtick pos=left,
ymin=7500, ymax=11000, 
ylabel={Area ($\upmu m^{2}$) }, 
y tick label style={ font=\fontsize{7}{7}\selectfont},
ylabel style={inner sep=0, font=\fontsize{7}{7}\selectfont}, 
ylabel shift=-2pt, ytickmin=7500,ytickmax=11000, 
legend columns=1, 
legend style={
font=\fontsize{7}{7}\selectfont,line width=0.5pt,
at={(0.55,0.81)}, anchor=center, 
/tikz/every even column/.append style={column sep=0.05cm},
draw=none, 
},
line width=0.55pt,
major tick length=2.5pt,
legend entries={{Retiming\&sizing},{VirtualSync+}},
] 
\addlegendimage{no markers, black,dashed}
\addlegendimage{no markers, blue}
\addplot[scatter, line width=0.5pt, black, dashed, dash pattern=on 3pt off 3pt,dash phase=8pt, domain = -5:5, 
scatter src=explicit symbolic]
coordinates {
(1,10474.8) [a]
(1.05,8710.4) [b]
(1.1,8775.1) [c]
(1.15,8531.7) [d]
(1.20,8451.4) [e]
(1.25,8126.3) [f]
};

\addplot[scatter, line width=0.5pt, blue, 
scatter src=explicit symbolic]
coordinates {
(0.97,9747.3) [a]
(1.03,8710.7) [b]
(1.08,8761.2) [c]
(1.13,8515.1) [d]
(1.18,8400.7) [e]
(1.23,8102.1) [f]
};

\end{axis}
\end{tikzpicture}
}
}
\\
\subfloat[][systemcaes]{\makebox{\pgfplotsset{compat=1.3,
    /pgfplots/ybar legend/.append style={ 
        /pgfplots/legend image code/.code={%
           \draw[##1,/tikz/.cd,yshift=-0.25em]
           (0cm,0cm) rectangle (5pt,0.8em);
        },
    }
}

\begin{tikzpicture}
\pgfplotstableread[row sep=\\]{
gap   re\\
1     37683.4\\
2     nan\\
3     26911\\
4     nan\\
5     21948.5\\
6     nan\\ 
7     21397.6\\
8     nan\\
9     21077.6\\
10    nan\\
11    20948.0\\
}\loadedtable

\pgfplotstableread[row sep=\\]{
gap    our\\     
1     36629.5\\
2     25233\\
3     nan\\
4     21142.6\\
5     nan\\
6     20999.5\\
7     nan\\
8     20973.1\\
9     nan\\
10    20950.1\\
11    nan\\
}\loadedtable

\begin{axis}[ 
scatter/classes={
a={mark=x,blue,solid},%
b={mark=triangle*,red,solid},%
c={mark=o,draw=black,solid},
d={mark=halfsquare*,green,solid},
e={mark=*,yellow,solid},
f={mark=star,purple,solid}},
xticklabels={0.97T, 1.026T, 1.082T,1.138T, 1.194T, 1.25T},
xtick={0.97, 1.026, 1.082,1.138, 1.194, 1.25},
xmin=0.97, xmax=1.25,
x=12.099cm, y=0.000486cm, 
x tick label style={rotate=0, xshift=0pt,yshift=-3pt,anchor=east, font=\fontsize{6.5}{6.5}\selectfont,  
inner sep=0}, 
xticklabel pos=left, xtick align=outside, xtick pos=left,
ymin=10000, ymax=15000, 
ylabel={Area ($\upmu m^{2}$) }, 
y tick label style={ font=\fontsize{7}{7}\selectfont},
ylabel style={inner sep=0, font=\fontsize{7}{7}\selectfont}, 
ylabel shift=-2pt, ytickmin=12000,ytickmax=15000, 
legend columns=1, 
legend style={
font=\fontsize{7}{7}\selectfont,line width=0.5pt,
at={(0.55,0.81)}, anchor=center, 
/tikz/every even column/.append style={column sep=0.05cm},
draw=none, 
},
line width=0.55pt,
major tick length=2.5pt,
legend entries={{Retiming\&sizing},{VirtualSync+}},
] 
\addlegendimage{no markers, black,dashed}
\addlegendimage{no markers, blue}
\addplot[scatter, line width=0.5pt, black, dashed, dash pattern=on 3pt off 3pt,dash phase=8pt, domain = -5:5, 
scatter src=explicit symbolic]
coordinates {
(1,14177.8) [a]
(1.05,11926.4) [b]
(1.1,11301.3) [c]
(1.15,11230.3) [d]
(1.20,10887.1) [e]
(1.25,10435.4) [f]
};

\addplot[scatter, line width=0.5pt, blue, 
scatter src=explicit symbolic]
coordinates {
(0.97,14120.1) [a]
(1.04,11800.1) [b]
(1.08,11150.1) [c]
(1.13,11120.1) [d]
(1.18,10771.7) [e]
(1.21,10499.1) [f]
};

\end{axis}
\end{tikzpicture}
}
}
\subfloat[][wb\_dma]{\makebox{\pgfplotsset{compat=1.3,
    /pgfplots/ybar legend/.append style={ 
        /pgfplots/legend image code/.code={%
           \draw[##1,/tikz/.cd,yshift=-0.25em]
           (0cm,0cm) rectangle (5pt,0.8em);
        },
    }
}

\begin{tikzpicture}
\pgfplotstableread[row sep=\\]{
gap   re\\
1     37683.4\\
2     nan\\
3     26911\\
4     nan\\
5     21948.5\\
6     nan\\ 
7     21397.6\\
8     nan\\
9     21077.6\\
10    nan\\
11    20948.0\\
}\loadedtable

\pgfplotstableread[row sep=\\]{
gap    our\\     
1     36629.5\\
2     25233\\
3     nan\\
4     21142.6\\
5     nan\\
6     20999.5\\
7     nan\\
8     20973.1\\
9     nan\\
10    20950.1\\
11    nan\\
}\loadedtable

\begin{axis}[ 
scatter/classes={
a={mark=x,blue,solid},%
b={mark=triangle*,red,solid},%
c={mark=o,draw=black,solid},
d={mark=halfsquare*,green,solid},
e={mark=*,yellow,solid},
f={mark=star,purple,solid}},
xticklabels={0.96T, 1.018T, 1.076T,1.134T, 1.192T, 1.25T},
xtick={0.96, 1.018, 1.076,1.134, 1.192, 1.25},
xmin=0.96, xmax=1.25,
x=12.099cm, y=0.0015780cm, 
x tick label style={rotate=0, xshift=0pt,yshift=-3pt,anchor=east, font=\fontsize{6.5}{6.5}\selectfont,  
inner sep=0}, 
xticklabel pos=left, xtick align=outside, xtick pos=left,
ymin=4800, ymax=6350, 
ylabel={Area ($\upmu m^{2}$) }, 
y tick label style={ font=\fontsize{7}{7}\selectfont},
scaled y ticks=base 10:-3,
ylabel style={inner sep=0, font=\fontsize{7}{7}\selectfont}, 
ylabel shift=-2pt, ytickmin=4800,ytickmax=6350, 
legend columns=1, 
legend style={
font=\fontsize{7}{7}\selectfont,line width=0.5pt,
at={(0.55,0.81)}, anchor=center, 
/tikz/every even column/.append style={column sep=0.05cm},
draw=none, 
},
line width=0.55pt,
major tick length=2.5pt,
legend entries={{Retiming\&sizing},{VirtualSync+}},
] 
\addlegendimage{no markers, black,dashed}
\addlegendimage{no markers, blue}
\addplot[scatter, line width=0.5pt, black, dashed, dash pattern=on 3pt off 3pt,dash phase=8pt, domain = -5:5, 
scatter src=explicit symbolic]
coordinates {
(1,5609.7) [a]
(1.05,5534.4) [b]
(1.1,5278.5) [c]
(1.15,5267.3) [d]
(1.20,5133.3) [e]
(1.25,5070.2) [f]
};

\addplot[scatter, line width=0.5pt, blue, 
scatter src=explicit symbolic]
coordinates {
(0.96,5973.0) [a]
(1.03,5724.3) [b]
(1.08,5252.4) [c]
(1.13,5234.3) [d]
(1.18,5121.6) [e]
(1.22,5050.6) [f]
};

\end{axis}
\end{tikzpicture}
}
}
\subfloat[][tv80]{\makebox{\pgfplotsset{compat=1.3,
    /pgfplots/ybar legend/.append style={ 
        /pgfplots/legend image code/.code={%
           \draw[##1,/tikz/.cd,yshift=-0.25em]
           (0cm,0cm) rectangle (5pt,0.8em);
        },
    }
}

\begin{tikzpicture}

\pgfplotstableread[row sep=\\]{
gap   re\\
1     37683.4\\
2     nan\\
3     26911\\
4     nan\\
5     21948.5\\
6     nan\\ 
7     21397.6\\
8     nan\\
9     21077.6\\
10    nan\\
11    20948.0\\
}\loadedtable

\pgfplotstableread[row sep=\\]{
gap    our\\     
1     36629.5\\
2     25233\\
3     nan\\
4     21142.6\\
5     nan\\
6     20999.5\\
7     nan\\
8     20973.1\\
9     nan\\
10    20950.1\\
11    nan\\
}\loadedtable

\begin{axis}[ 
scatter/classes={
a={mark=x,blue,solid},%
b={mark=triangle*,red,solid},%
c={mark=o,draw=black,solid},
d={mark=halfsquare*,green,solid},
e={mark=*,yellow,solid},
f={mark=star,purple,solid}},
xticklabels={1T, 1.05T, 1.10T,1.15T, 1.20T, 1.25T},
xtick={1, 1.05, 1.10,1.15, 1.20, 1.25},
xmin=1, xmax=1.25,
x=14.160cm, y=0.000490cm, 
x tick label style={rotate=0, xshift=0pt,yshift=-3pt,anchor=east, font=\fontsize{6.5}{6.5}\selectfont,
inner sep=0}, 
xticklabel pos=left, xtick align=outside, xtick pos=left,
ymin=9000, ymax=14000, 
ylabel={Area ($\upmu m^{2}$) }, 
y tick label style={ font=\fontsize{7}{7}\selectfont},
ylabel style={inner sep=0, font=\fontsize{7}{7}\selectfont}, 
ylabel shift=-2pt, ytickmin=9000,ytickmax=14000, 
legend columns=1, 
legend style={
at={(0.53,0.81)}, anchor=center, 
/tikz/every even column/.append style={column sep=0.05cm},
draw=none, 
},
line width=0.55pt,
major tick length=2.5pt,
legend style={font=\fontsize{7}{7}\selectfont},
legend entries={{Retiming\&sizing},{VirtualSync+}},
] 
\addlegendimage{no markers, black,dashed}
\addlegendimage{no markers, blue}
\addplot[scatter, line width=0.5pt, black, dashed, dash pattern=on 3pt off 3pt,dash phase=8pt, domain = -5:5, 
scatter src=explicit symbolic]
coordinates {
(1,11692.8) [a]
(1.05,11534.0) [b]
(1.1,10020.0) [c]
(1.15,9786.1) [d]
(1.20,9551.0) [e]
(1.25,9713.8) [f]
};

\addplot[scatter, line width=0.5pt, blue, 
scatter src=explicit symbolic]
coordinates {
(1,11465.7) [a]
(1.04,10953.1) [b]
(1.09,10010.9) [c]
(1.14,9698.4) [d]
(1.19,9410.8) [e]
(1.23,9655.7) [f]
};

\end{axis}
\end{tikzpicture}
}
}
\subfloat[][systemcdes]{\makebox{\pgfplotsset{compat=1.3,
    /pgfplots/ybar legend/.append style={ 
        /pgfplots/legend image code/.code={%
           \draw[##1,/tikz/.cd,yshift=-0.25em]
           (0cm,0cm) rectangle (5pt,0.8em);
        },
    }
}

\begin{tikzpicture}
\pgfplotstableread[row sep=\\]{
gap   re\\
1     37683.4\\
2     nan\\
3     26911\\
4     nan\\
5     21948.5\\
6     nan\\ 
7     21397.6\\
8     nan\\
9     21077.6\\
10    nan\\
11    20948.0\\
}\loadedtable

\pgfplotstableread[row sep=\\]{
gap    our\\     
1     36629.5\\
2     25233\\
3     nan\\
4     21142.6\\
5     nan\\
6     20999.5\\
7     nan\\
8     20973.1\\
9     nan\\
10   20950.1\\
11   nan\\
}\loadedtable

\begin{axis}[ 
scatter/classes={
a={mark=x,blue,solid},%
b={mark=triangle*,red,solid},%
c={mark=o,draw=black,solid},
d={mark=halfsquare*,green,solid},
e={mark=*,yellow,solid},
f={mark=star,purple,solid}},
xticklabels={0.99T, 1.042T, 1.094T,1.146T, 1.198T, 1.25T},
xtick={0.99, 1.042, 1.094,1.146, 1.198, 1.25},
xmin=0.99, xmax=1.25,
x=13.562cm, y=0.000977cm, 
x tick label style={rotate=0, xshift=0pt,yshift=-3pt,anchor=east, font=\fontsize{6.5}{6.5}\selectfont,  
inner sep=0}, 
xticklabel pos=left, xtick align=outside, xtick pos=left,
ymin=4000, ymax=6500, 
ylabel={Area ($\upmu m^{2}$) }, 
y tick label style={ font=\fontsize{7}{7}\selectfont, 
 /pgf/number format/.cd,
        fixed,
        fixed zerofill,
        precision=1,
        /tikz/.cd
},
scaled y ticks=base 10:-3,
ylabel style={inner sep=0, font=\fontsize{7}{7}\selectfont}, 
ylabel shift=-2pt, ytickmin=4300,ytickmax=6550, 
legend columns=1, 
legend style={
font=\fontsize{7}{7}\selectfont,line width=0.5pt,
at={(0.55,0.81)}, anchor=center, 
/tikz/every even column/.append style={column sep=0.05cm},
draw=none, 
},
line width=0.55pt,
major tick length=2.5pt,
legend entries={{Retiming\&sizing},{VirtualSync+}},
] 
\addlegendimage{no markers, black,dashed}
\addlegendimage{no markers, blue}
\addplot[scatter, line width=0.5pt, black, dashed, dash pattern=on 3pt off 3pt,dash phase=8pt, domain = -5:5, 
scatter src=explicit symbolic]
coordinates {
(1,5500.9) [a]
(1.05,5488.7) [b]
(1.1,5028.7) [c]
(1.15,4751.8) [d]
(1.20,4827.6) [e]
(1.25,4420.9) [f]
};

\addplot[scatter, line width=0.5pt, blue, 
scatter src=explicit symbolic]
coordinates {
(0.99,5553.8) [a]
(1.03,5512.5) [b]
(1.08,5085.1) [c]
(1.14,4769.1) [d]
(1.18,4910.2) [e]
(1.23,4445.8) [f]
};

\end{axis}
\end{tikzpicture}
}
}
\caption{Area comparison between relaxed retiming and sizing and \titlenym\ when the target clock period is relaxed to 1.05T, 1.10T, 1.15T, 1.20T and 1.25T, where T is the clock period achieved by extreme R\&S.}
\label{fig:area_per}%
\end{minipage}

\end{figure*}
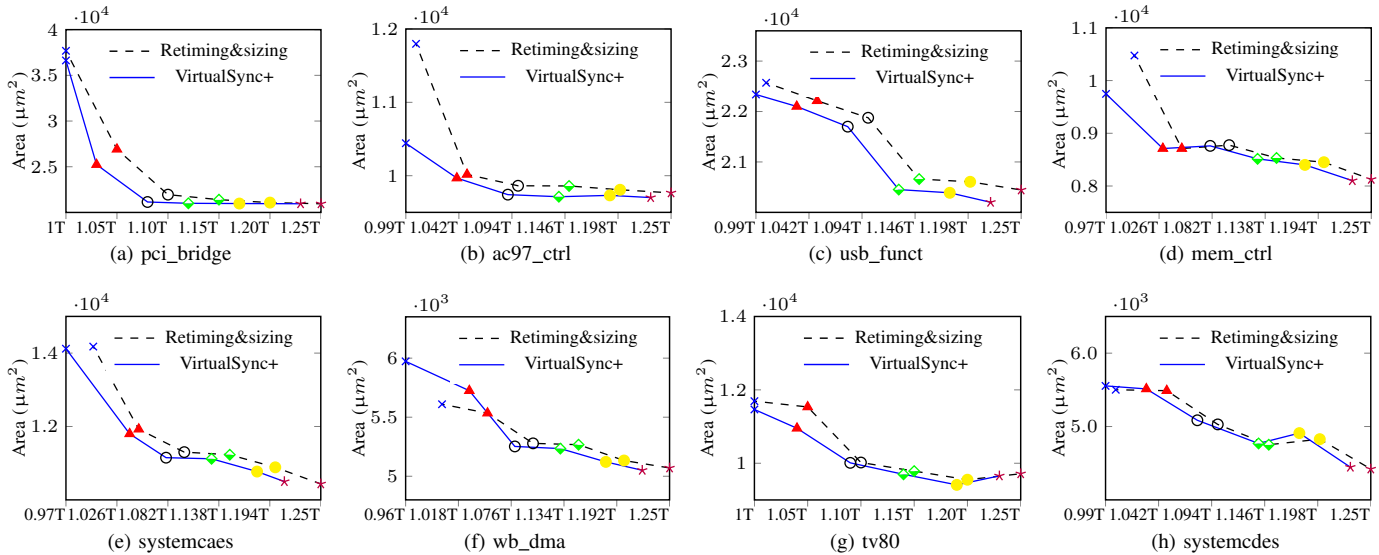

The column $\boldsymbol{n_{t}}$ in Table~\ref{tb_test} shows the final clock period 
reduction compared with the circuits after extreme R\&S. 
 The maximum and average reduction are 8.5\% and 3.2\%, respectively, which 
resulted from the compensation between flip-flop stages and the removal
of clock-to-q delays and setup time requirements on critical paths.  
For all the cases, the minimum clock periods have been pushed even further than
those from extreme R\&S. %
The timing performance improvement $\boldsymbol{n_{t}}$ is achieved with relaxed \titlenym\ described in Section~\ref{sec:impl}.  
This method can be used to quickly evaluate the benefit of \titlenym\ in circuit design.

The area increase of the proposed method compared with extreme R\&S 
is shown in column $\boldsymbol{n_{a}}$.  
In the cases with area increase, the overhead is still negligible; in other
cases, the area is even smaller because unnecessary flip-flops were removed 
in the proposed framework. 
The last column $t_r$ in Table~\ref{tb_test} shows the runtime of the proposed method. 
Since the ILP formulation with the complete model in Section~\ref{sec:mode_delay_unit}  
is NP-hard, it is impractical to find a solution with respect to area and clock period. In the experiments, 
the runtime with iterative relaxations 
is acceptable in remedying remaining timing violations for late design stage.  %

In the proposed framework, sequential delay units are first inserted only at
necessary locations to delay signal propagation. Afterwards,
more of them are used to replace buffers to reduce area, as described in 
Section~\ref{sec:impl}.  
To demonstrate the effectiveness of the buffer replacement with sequential units 
in the proposed framework, 
we compare the numbers of sequential delay units without 
and with buffer replacement, as shown in 
Figure~\ref{fig:se_cmp}. %
This figure shows an increase in the number of
such delay units replacing buffers. %
Figure~\ref{fig:buffer_replace} shows 
the ratio of the chip area after buffer replacement to that before buffer replacement.
In all test cases, 
the area after buffer replacement is smaller than that before buffer replacement, 
demonstrating the efficiency of
sequential delay units in delaying fast signals.

The comparison of the area overhead, shown as $\boldsymbol{n_{a}}$ in Table~\ref{tb_test}, is between  
the clock period achieved by extreme R\&S and the 
clock period reduced further by the proposed method. To demonstrate the area efficiency of the proposed method, we
also compared the proposed method and the extreme R\&S with the same clock period from the latter.  
The results are shown in \figname~\ref{fig:samet_area_cmp}.
In most cases, our
framework can further reduce the area achieved by extreme R\&S.  

\subsection{Experimental Results with Circuit Fine-Tuning in Design Compiler}

The experimental results described in Section~\ref{sec:secondstep} are generated with the timing 
optimization using relaxed \titlenym\ in Section~\ref{sec:impl}. 
These intermediate results 
might be inconsistent with commercial tools due to the simplified model. 
To provide more consistent results,  
we fine-tune the optimized circuits with \titlenym\ in Design Compiler, as described in 
Section~\ref{sec:integration}. 
After this fine-tuning, we reevaluated the improvement of 
timing performance and area reduction with respect to the circuits after extreme R\&S.
The comparisons between the clock period reduction and the area reduction with and without the fine-tuning 
with Design Compiler are illustrated 
in \figname~\ref{fig:timing_com} and \figname~\ref{fig:area_com}, respectively. 
According to these two figures, 
it is clear that in most cases the timing performance with the fine-tuning %
can still be pushed beyond that after extreme R\&S while the area is reduced. 
Specifically, the timing 
performance can be improved by up to 4\% (average 1.5\%) compared with that
after extreme R\&S. 
However, the timing performance improvement with the fine-tuning is less than that without the fine-tuning in most cases.  
\reviewhl{One of the reasons is that 
the commercial 
tools do not support 
the timing optimization and analysis of wave-pipelining paths where latches are inserted to block fast signals.
}
Accordingly, only flip-flops are used as delay units to block fast signals in the fine-tuning of \titlenym\ with Design Compiler.  
Another reason of this phenomenon is that  
in the iterative relaxation without the fine-tuning, gates are sized ideally 
without  
considering the input transitions and output loads of gates. 
Therefore, the timing performance improvement without the fine-tuning is 
a theoretical upper bound of timing performance improvement, 
which can be used to quickly evaluate the benefit of \titlenym\ in circuit design. 
If it is beneficial to apply \titlenym, designers can then adopt the proposed fine-tuning method with Design Compiler
to generate the optimized circuits.
With \titlenym, 
the power consumption and power-delay-area product of the optimized circuits
can also be reduced %
in most cases. The comparison is illustrated in \figname~\ref{fig:power_com}.

In the experiments, 
the extreme R\&S to push the timing performance 
beyond the limit of traditional sequential designs 
is realized by iteratively reducing the clock period and optimizing 
the circuits with retiming and sizing in Design Compiler until the circuits cannot be optimized further anymore.  
This method squeezes the timing performance 
at the cost of area overhead. %
In cases where timing performance does not need to be pushed so far for the sake of area, 
\titlenym\ can still outperform the method combining retiming and sizing in terms of area and speed. 
To demonstrate the effectiveness of \titlenym\ in such cases, 
the clock period achieved by extreme R\&S, denoted as $T$, is relaxed, denoted as relaxed R\&S. 
For 
example, we used the relaxed clock periods 1.05$T$, 1.10$T$, 1.15$T$, 1.20$T$ and 1.25$T$ 
to compare 
\titlenym\ and relaxed R\&S
in terms of area and timing performance.  
With these relaxed clock periods, 
\titlenym\ with fine-tuning in commercial tools is applied to 
improve the timing performance of the circuits optimized with relaxed R\&S. 
The comparison in timing performance and area with respect to the relaxed clock periods 
is illustrated in \figname~\ref{fig:area_per}, where each pair of symbols, e.g., triangles, represent a comparison in 
timing and area between \titlenym\ and relaxed R\&S with a specified clock period.  
For example, in \figname~\ref{fig:area_per}(a), the pair of triangles represent the results 
with \titlenym\ and relaxed R\&S when the  
clock period is 1.05$T$. Since the result with \titlenym\ is 
at the lower left of the result with the relaxed R\&S, %
the timing performance with \titlenym\ is better than that of relaxed R\&S, while the area overhead is reduced. 
By connecting these symbols together,  %
we can see that in most cases \titlenym\ achieves a smaller area 
under the same clock period 
compared with 
relaxed R\&S.
In addition, 
\titlenym\ achieves a smaller clock period under the same area cost compared with  
relaxed R\&S.

\section{Conclusion}\label{sec:conclusion}
In this paper, we have proposed a new timing model, \titlenym, in which 
sequential components and combinational logic gates are considered as delay units.
They provide different delay effects on signal propagations on short and long paths. 
With this new timing model, a timing optimization framework has been proposed
to insert delay units only at necessary locations.
In addition, we further enhance the optimization with \titlenym\ by fine-tuning with
commercial design tools,
e.g., Design Compiler from
Synopsys, to
achieve more accurate result.
Experimental results show that circuit
performance can be improved by up to 4\% (average 1.5\%) compared with that
after extreme retiming and sizing, while the increase of area is
still negligible. 
\reviewhl{In the future work, 
we will %
enhance the framework to 
integrate \titlenym\ with physical design tools to fully automate virtual synchronization 
into the EDA flow. %
To implement wave-pipelining paths with latches using the commercial tools, 
however, either the commercial timing tools provide more flexible timing constraints for these paths or
they allow access to the APIs.} %

\let\oldthebibliography=\thebibliography
\let\endoldthebibliography=\endthebibliography
\renewenvironment{thebibliography}[1]{%
\begin{oldthebibliography}{#1}%
\setlength{\itemsep}{0.8ex}%
\def\baselinestretch{4}
\setlength{\baselineskip}{\baselinestretch\mylength}
\fontsize{6.4pt}{1}\selectfont
}%
{%
\end{oldthebibliography}%
}

\bibliographystyle{IEEEtran}

\begin{footnotesize}
\vskip -40pt
\begin{IEEEbiography}
[{\includegraphics[width=1in,height=1.25in,clip,keepaspectratio]{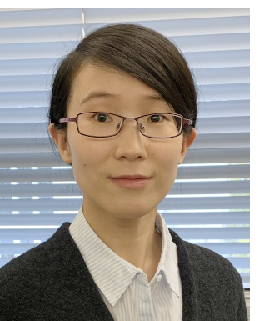}}] 
{Grace Li Zhang} 
received the Dr.-Ing. degree from the Technical University of Munich (TUM), Munich, Germany, in 2018. She is currently a postdoctoral researcher pursuing Habilitation at the Chair of Electronic Design Automation, TUM, where she leads the research team on heterogeneous computing. Her research interests include neural networks and neuromorphic computing, computer architectures, and machine learning for EDA. She has served/is serving on the technical committee of several conferences including DAC, ICCAD, ASP-DAC, GLSVLSI etc. 
\end{IEEEbiography}

\vskip -25pt
\begin{IEEEbiography}
[{\includegraphics[width=1in,height=1.25in,clip,keepaspectratio]{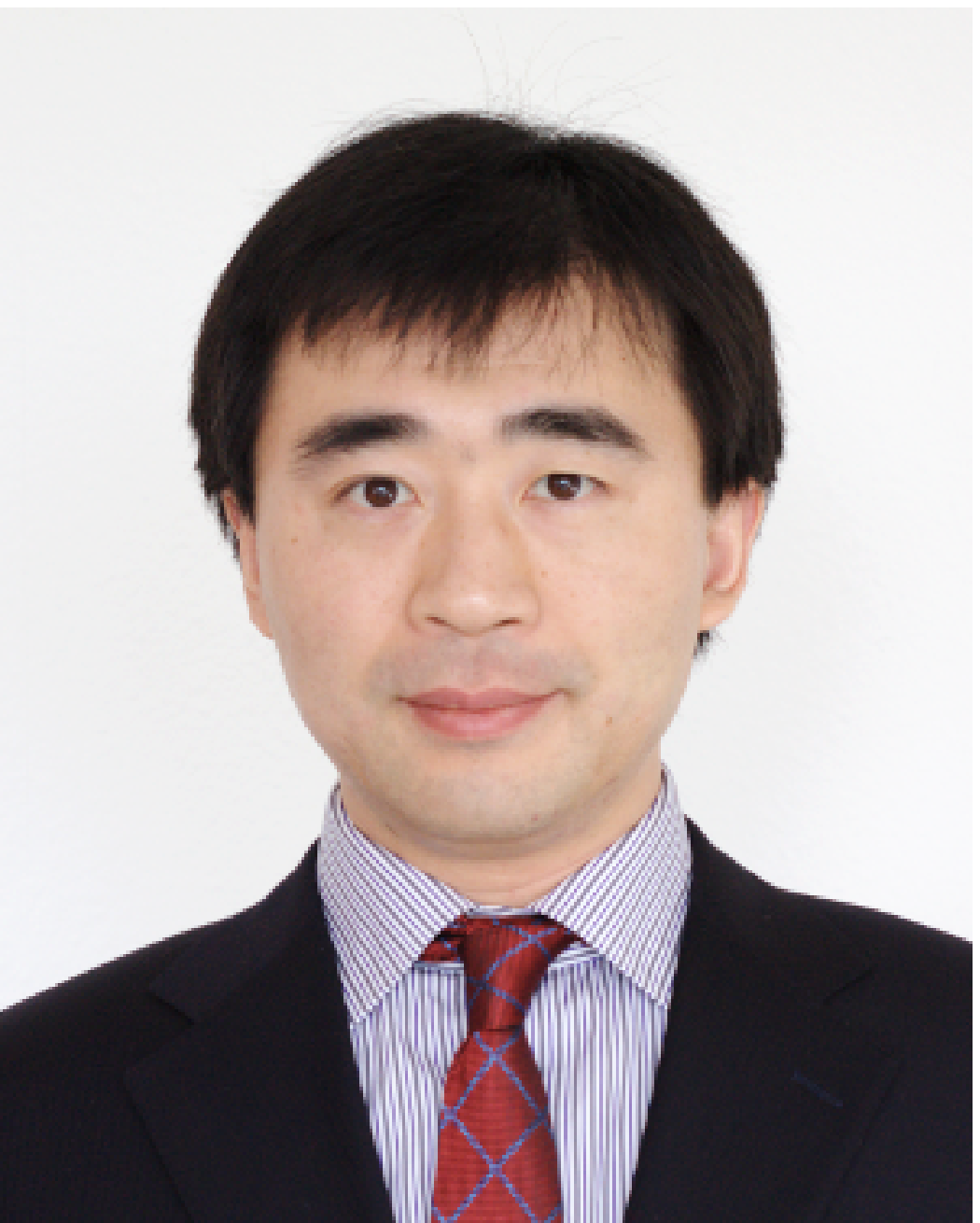}}] 
{Bing Li} received the Dr.-Ing. degree from Technical University of Munich (TUM), Munich, Germany, in 2010 and finished the Habilitation there in 2018. He is currently a researcher with the Chair of Electronic Design Automation, TUM. His current research interests include high-performance and low-power design, AI hardware, and emerging systems. He has served on the Technical Program Committees of several conferences including DAC, ICCAD, DATE, ASP-DAC, etc. 
\end{IEEEbiography}

\vskip -20pt
\begin{IEEEbiography}
[{\includegraphics[width=1in,height=1.25in,clip,keepaspectratio]{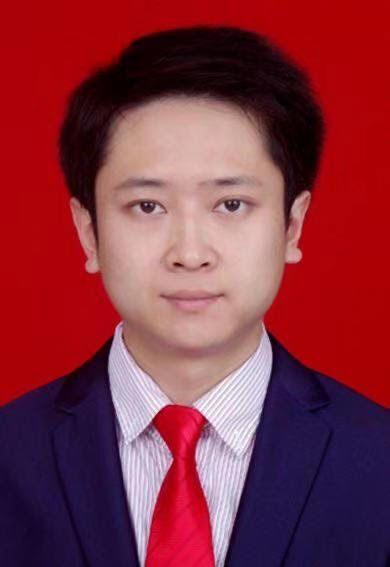}}] 
{Xing Huang} received the Ph.D. degree in electronic science and technology from Fuzhou University, Fuzhou, China, in 2018. 
He is currently a Postdoctoral Research Fellow with the Chair of Electronic Design Automation, Technical University of Munich, Germany, sponsored by the Alexander von Humboldt Foundation.  His current research interests include design automation for microfluidic biochips and integrated circuits.
\end{IEEEbiography}

\vskip -20pt
\begin{IEEEbiography}
[{\includegraphics[width=1in,height=1.25in,clip,keepaspectratio]{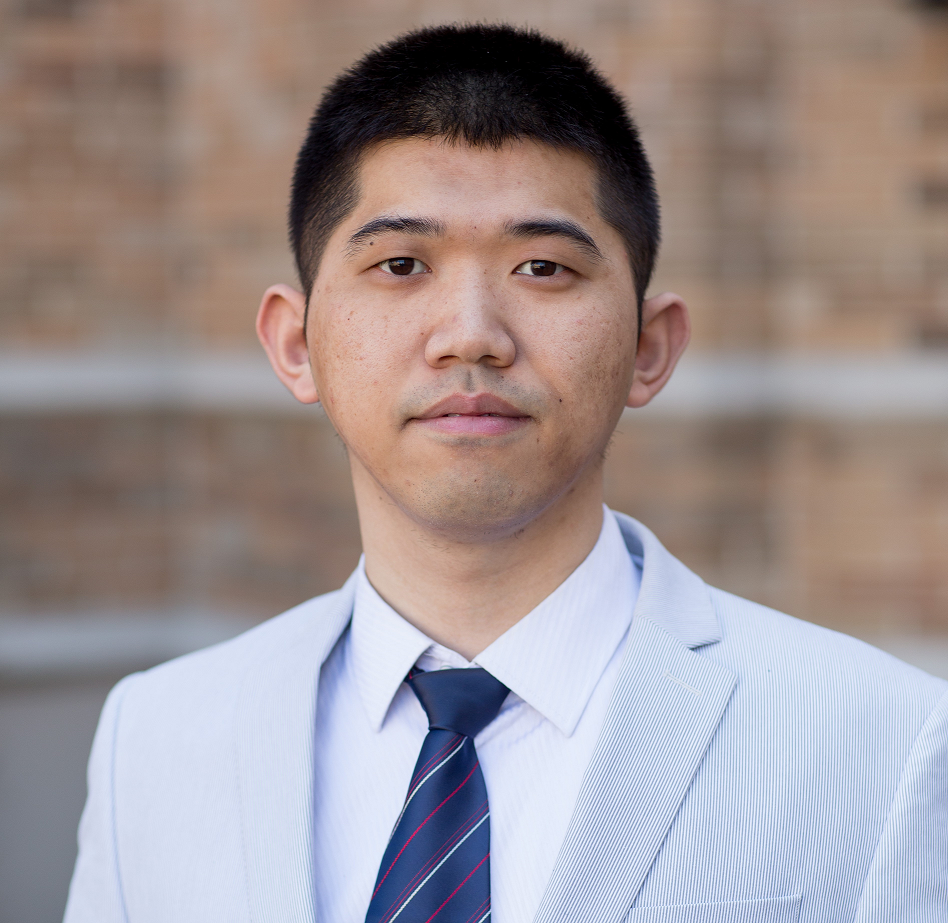}}]
{Xunzhao Yin} is an assistant professor of the College of Information Science and Electronic Engineering at Zhejiang University. He received his Ph.D. degree from University of Notre Dame in 2019 and B.S. degree from Tsinghua University in 2013, respectively.
His research focuses on emerging circuit/architecture designs and novel computing paradigms with both CMOS and emerging technologies. 
\end{IEEEbiography}

\vskip -20pt
\begin{IEEEbiography}
[{\includegraphics[width=1in,height=1.25in,clip,keepaspectratio]{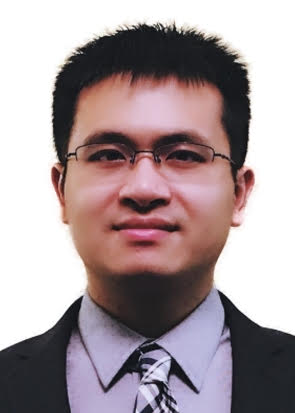}}]
{Cheng Zhuo} 
(M'12--SM'16)
received the B.S. and M.S. degrees in electronic engineering from Zhejiang University, Hangzhou, China, in 2005 and 2007, respectively. He received the Ph.D. degree in computer science \& engineering from the University of Michigan, Ann Arbor, MI, USA, in 2010. He is currently with Zhejiang University as a Professor in the college of Information Science \& Electronic Engineering. His current research interests include computing in memory, deep learning, and general VLSI EDA areas. 
\end{IEEEbiography}

\vskip -15pt
\begin{IEEEbiography}
[{\includegraphics[width=1in,height=1.25in,clip,keepaspectratio]{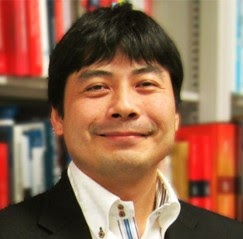}}]
{Masanori Hashimoto} received the Ph.D. degrees
from Kyoto University in 2001. He is currently a Professor with the Graduate
School of Informatics, Kyoto University. His current research interests include
design for reliability, timing and power integrity analysis, reconfigurable
computing, soft error characterization, and low-power circuit design.
\end{IEEEbiography}
\vskip -15pt

\begin{IEEEbiography}
[{\includegraphics[width=1in,height=1.25in,clip,keepaspectratio]{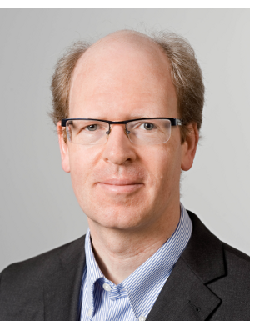}}] 
{Ulf Schlichtmann} %
 received the Dipl.-Ing. and Dr.-Ing. Degrees in electrical engineering and information technology from Technical University of Munich (TUM), Munich, Germany, in 1990 and 1995, respectively. He was with Siemens AG, Munich, and Infineon Technologies AG, Munich, from 1994 to 2003, where he held various technical and management positions in design automation, design libraries, IP reuse, and product development. Since 2003, he is Professor and the Head of the Chair of Electronic Design Automation at TUM. His current research interests include computer-aided design of electronic circuits and systems, with an emphasis on designing reliable and robust systems. Increasingly, he focuses on emerging technologies such as lab-on-a-chip and photonics.
\end{IEEEbiography}
\end{footnotesize}
\vfill

\end{document}